\pdfoutput=1
\RequirePackage{ifpdf}
\ifpdf 
\documentclass[pdftex]{sigma}
\else
\documentclass{sigma}
\fi

\DeclareMathOperator{\minor}{minor}

\theoremstyle{plain}
\newtheorem{cnj}[theorem]{Conjecture}

\theoremstyle{definition}
\newtheorem{cnv}{Convention}

\begin{document}

\allowdisplaybreaks

\renewcommand{\PaperNumber}{117}

\FirstPageHeading

\ShortArticleName{Relations in Grassmann Algebra Corresponding to Pachner Moves}

\ArticleName{Relations in Grassmann Algebra Corresponding\\ to Three- and Four-Dimensional Pachner Moves}

\Author{Igor G.~KOREPANOV}

\AuthorNameForHeading{I.G.~Korepanov}

\Address{Moscow State University of Instrument Engineering and Computer Sciences,\\
 20 Stromynka Str.,  Moscow 107996, Russia}
\Email{\href{mailto:paloff@ya.ru}{paloff@ya.ru}}

\ArticleDates{Received May 15, 2011, in f\/inal form December 16, 2011; Published online December 18, 2011}

\Abstract{New algebraic relations are presented, involving anticommuting Grassmann variables and Berezin integral, and corresponding naturally to Pachner moves in three and four dimensions. These relations have been found experimentally~-- using symbolic computer calculations; their essential new feature is that, although they can be treated as deformations of relations corresponding to torsions of acyclic complexes, they can no longer be explained in such terms. In the simpler case of three dimensions, we def\/ine an invariant, based on our relations, of a piecewise-linear manifold with triangulated boundary, and present example calculations conf\/irming its nontriviality.}

\Keywords{Pachner moves; Grassmann algebras; algebraic topology}

\Classification{15A75; 55-04; 57M27, 57Q10; 57R56}

\section{Introduction}\label{s:I}

The main aim of this paper is to present some results of symbolic calculations, namely, new algebraic relations with anticommuting Grassmann variables and Berezin integral, corresponding naturally to Pachner moves in three and four dimensions. These results do not rely on any f\/inished theory; they were found starting from considerations related to some unusual chain complexes introduced in our previous works, an then using just free search on a computer, some heuristics like parameter counting, some hardly explainable tricks, and a hope that interesting relations may exist. Our software included GAP~\cite{GAP}, our own package PL~\cite{PL}, and Maxima~\cite{maxima}.

The essential new feature of our algebraic relations is that, although they can be treated as deformations of already known relations corresponding to torsions of acyclic complexes, they can no longer be formulated and explained in such terms\footnote{Note also that in our previous works, the (simpler) relations corresponding to Pachner moves were derived using direct calculations as well, using only some ``partial'' theoretical considerations.}.

As we have said, our calculations~-- computer experiments~-- belong to three- and, what seems the most interesting, four-dimensional topology. As the four-dimensional case is more complicated, we restrict ourself to presenting relations corresponding to Pachner moves $3\to 3$ and $2\to 4$ (the f\/irst of them proved in full on a computer, while the second is a conjecture conf\/irmed by \emph{numerical}\footnote{As opposed to \emph{symbolic}.} calculations), and a (conjectured) formula for a ``partial'' manifold invariant~-- a Grassmann algebra element preserved by \emph{these} moves, leaving for future work both the construction of a ``full'' invariant and its calculations for specif\/ic manifolds. In contrast with this, we def\/ine an actual invariant of a \emph{three}-dimensional piecewise-linear manifold with triangulated boundary, based on our relations, and present example calculations conf\/irming its nontriviality.

This introduction continues with a brief reminder of what Pachner moves and Grassmann algebras are, in Subsections \ref{ss:P} and~\ref{ss:B} respectively. Then, we explain the organization of the (main part of the) paper in Subsection~\ref{ss:org}.

\subsection{Pachner moves and algebraic relations}\label{ss:P}

Pachner moves~\cite{P} (see also a very good educational text~\cite{L}) are elementary rebuildings of a~manifold triangulation. Their importance is due to the fact that any triangulation of a closed piecewise-linear (PL) manifold can be transformed into any other triangulation by a sequence of these moves, of which, moreover, there exists only a f\/inite number in any manifold dimension. Thus, if we have algebraic formulas whose structure corresponds, in some natural sense, to all Pachner moves in a given dimension, then there is a big hope that we will be able~-- maybe with the help of some additional tools~-- to build a manifold invariant based on such formulas. Also, some experience shows that if there is a formula corresponding to just \emph{one} Pachner move, then it makes a strong sense to search for more formulas for other moves.

Some modif\/ications are to be done if we consider manifolds with boundary. Pachner proves that, in this case, any triangulation can be transformed into any other triangulation by a sequence of \emph{shellings} and \emph{inverse shellings}. Each of these operations af\/fects the boundary and changes its triangulation. If we want, however, to construct an invariant of a manifold with a \emph{fixed boundary triangulation}~-- and this is what we do in Section~\ref{s:di} of the present paper~-- we must choose another way. Namely, we consider \emph{relative} Pachner moves, that is, moves not changing the boundary triangulation. The resulting invariant depends thus on the way the manifold is glued to its boundary\footnote{Recall that, for instance, in three dimensions there are many ways to glue a f\/illed pretzel to its boundary.}.

\subsubsection{Pachner moves in three dimensions}\label{sss:P3}

There are four Pachner moves in three dimensions: $2\leftrightarrow 3$ and $1\leftrightarrow 4$.

\begin{figure}[t]
\centering
\includegraphics[scale=0.4]{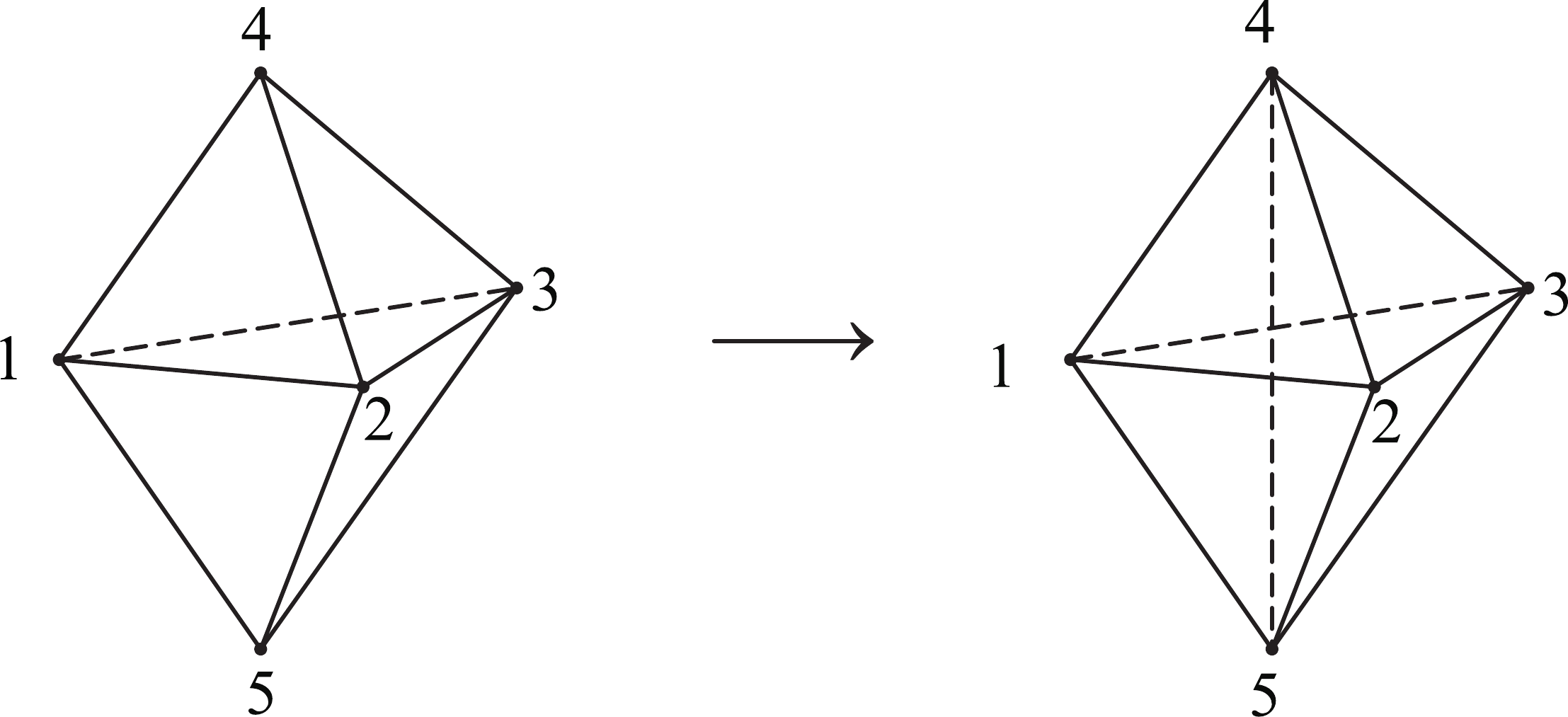}
\caption{Pachner move $2\to 3$.}
\label{f:23}
\end{figure}

Pachner move $2\to 3$ is an elementary rebuilding of a 3-manifold triangulation, which replaces two adjacent tetrahedra $(1234)$ and~$(1235)$ (where of course $(1234)$ is the tetrahedron with vertices $1$, $2$, $3$ and~$4$, and so on) with three tetrahedra $(1245)$, $(1345)$ and~$(2345)$ occupying the same place in the manifold, as in Fig.~\ref{f:23}.

\begin{figure}[t]
\centering
\includegraphics[scale=0.4]{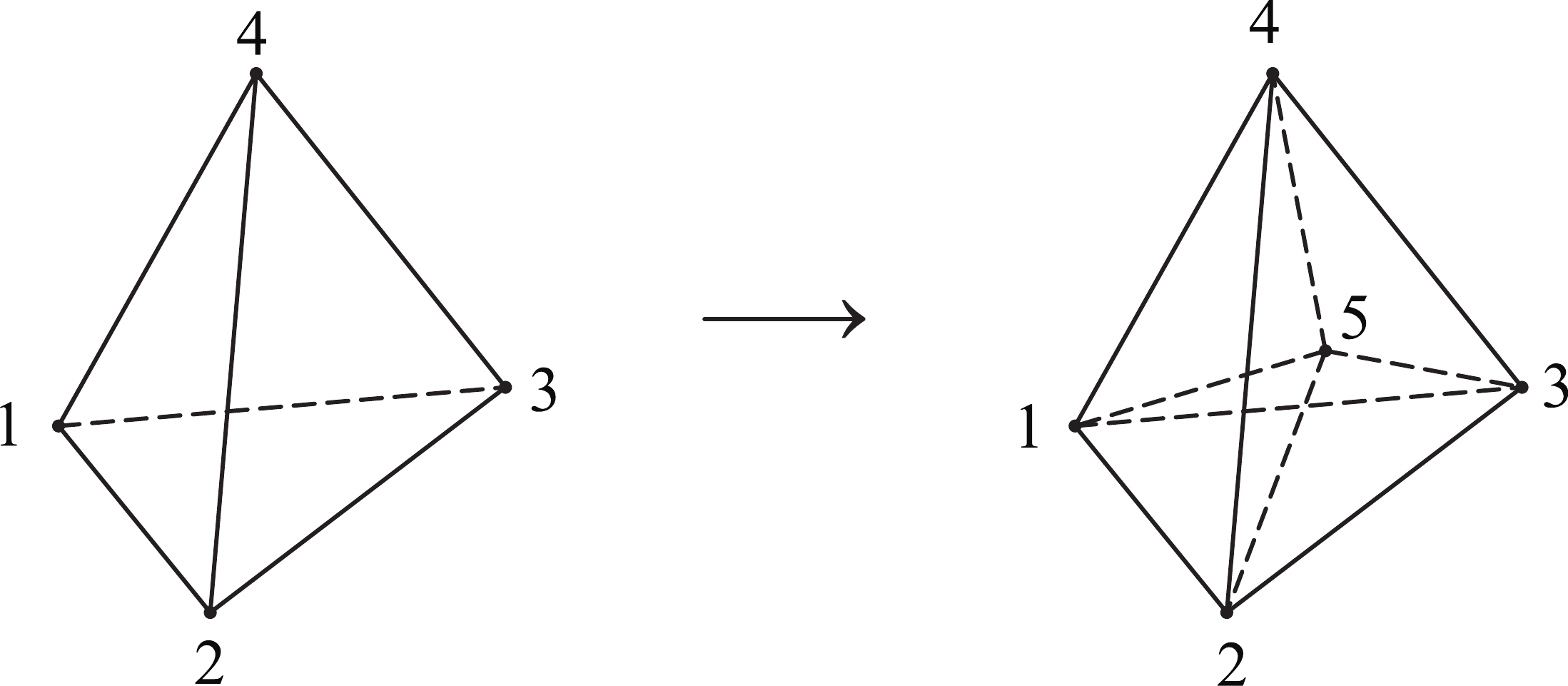}
\caption{Pachner move $1\to 4$.}
\label{f:14}
\end{figure}

Pachner move $1\to 4$ adds a new vertex~$5$ inside a tetrahedron~$(1234)$ and replaces it with tetrahedra $(1235)$, $(1245)$, $(1345)$ and~$(2345)$, see Fig.~\ref{f:14}.

Two other moves are their inverses.

\begin{remark}
Strictly speaking, our triangulations are not exactly like in~\cite{L}: we allow using dif\/ferent simplices having the same boundary components, see Fig.~\ref{f:1a} below for a good example. Such triangulations are sometimes called ``noncombinatorial''~-- a not very appropriate term, because combinatorics is exactly what we widely use, in particular, in our computer package~PL~\cite{PL}. Nevertheless, all our operations can be quite easily translated into the language of~\cite{L}, see, for instance, again~\cite[Section~2]{dkm}.
\end{remark}

\subsubsection{Pachner moves in four dimensions}\label{sss:P4}

There are f\/ive Pachner moves in four dimensions: $3\to 3$, $2\leftrightarrow 4$ and $1\leftrightarrow 5$.

In Section~\ref{s:4d}, we will be dealing with Pachner moves $3\to 3$ and $2\to 4$, their descriptions are given there in Subsections \ref{sss:33} and~\ref{sss:24} respectively.

In this paper, we do not present formulas for a move $1\to 5$, leaving this for further work. We would like only to explain here that this move consists in adding a new vertex~$6$ inside a given four-simplex~$(12345)$ and joining this new vertex with the boundary of~$(12345)$. This leads to the latter being divided into f\/ive tetrahedra $(12346)$, $(12356)$, $(12456)$, $(13456)$, and~$(23456)$.

\subsection{Grassmann algebras and Berezin integral}\label{ss:B}

A \emph{Grassmann algebra} over a f\/ield~$\mathbb F$~-- for which we can take in this paper any f\/ield of characte\-ris\-tic${}\ne 2$~-- is an associative algebra with unity, having generators~$a_i$ and relations
\begin{equation}\label{aiaj}
a_i a_j = -a_j a_i .
\end{equation}
As this implies for $i=j$ that $a_i^2 =0$, any element of a Grassmann algebra is a polynomial of degree $\le 1$ in each~$a_i$. For a given Grassmann monomial, by its degree we understand its total degree in all Grassmann variables; if an element of Grassmann algebra includes only monomials of odd degrees, it is called odd; if it includes only monomials of even degrees, it is called even.

The \emph{exponent} is def\/ined by the standard Taylor series. For example,
\[
\exp (a_1a_2) = 1+a_1a_2 .
\]
If at least one of $\varphi_1$ and~$\varphi_2$ is even, then
\begin{equation}\label{ee}
\exp(\varphi_1) \exp(\varphi_2)=\exp(\varphi_1+\varphi_2).
\end{equation}

The \emph{Berezin integral}~\cite{B} is an $\mathbb F$-linear operator in a Grassmann algebra def\/ined by equalities
\begin{equation}\label{iB}
\int \mathrm da_i =0, \qquad \int a_i\, \mathrm da_i =1, \qquad \int gh\, \mathrm da_i = g \int h\, \mathrm da_i,
\end{equation}
if $g$ does not depend on~$a_i$ (that is, generator~$a_i$ does not enter the expression for~$g$); multiple integral is understood as iterated one, according to the following model:
\begin{equation}\label{mB}
\iint ab\, \mathrm db\, \mathrm da = \int a \left( \int b\, \mathrm db \right) \mathrm da = 1 .
\end{equation}

The \emph{left derivative}~$\partial/\partial a$ w.r.t.\ a Grassmann generator~$a$ for a monomial~$f$ is def\/ined as follows: if $f$ does not contain~$a$, then $\partial f/\partial a = 0$, otherwise bring $a$ to the left using commutation relations~\eqref{aiaj} and strike it out.

\begin{remark}
A curious feature of Grassmann--Berezin calculus is that the integral is the same operation as the right derivative (def\/ined in obvious analogy with the left one). Nevertheless, using the two names for one operation makes sense because sometimes this is an analogue of the usual ``commutative'' integral, and sometimes~-- of the derivative. Moreover, it may turn out that this feature makes Grassmann--Berezin calculus a powerful tool in constructing unusual algabraic structures.
\end{remark}

\subsection{Organization of the paper}\label{ss:org}

A ``theorem'' in this paper is a statement proved either in a traditional way, or using a computer and software for symbolic calculations. A ``conjecture'' is a statement whose correctness raises practically no doubt but which has not been strictly proved. For instance, this can be a formula whose validity has been checked for some arbitrarily chosen set(s) of values of indeterminates, while checking it without assigning numerical values to indeterminates was beyond the available computational powers. Theorems and conjectures are numbered consecutively; ``partial verif\/ication'' of a conjecture corresponds to the ``proof'' of a theorem.

Below:
\begin{itemize}\itemsep=0pt
\item in Section~\ref{s:3d}, we present our Grassmann algebraic relations corresponding to Pachner moves in three dimensions,
\item in Section~\ref{s:4d}, we present similar relations corresponding to Pachner moves $3\to 3$ and $2\leftrightarrow 4$ in four dimensions,
\item in Section~\ref{s:di}, we return to the three-dimensional case and consider an example of invariant of three-manifolds with triangulated boundary, showing its nontriviality,
\item in Section~\ref{s:D}, we brief\/ly discuss our results and further research.
\end{itemize}

\section[Three dimensions: relations corresponding to moves $2\leftrightarrow 3$ and $1\leftrightarrow 4$]{Three dimensions: relations corresponding\\ to moves $\boldsymbol{2\leftrightarrow 3}$ and $\boldsymbol{1\leftrightarrow 4}$}\label{s:3d}

\subsection{Recalling the ``undeformed'' relations}\label{ss:C}

\subsubsection{The chain complex}\label{sss:dcc}

The starting point for the invariants~-- f\/ield theory amplitudes~-- introduced in this paper, is a particular~-- ``scalar''~-- case of the theory exposed in paper~\cite{bk}. Namely, we begin with the following chain complex built for a triangulated orientable three-manifold~$M$ with boundary:
\begin{equation}\label{c}
0 \longrightarrow \mathbb C^{N'_0} \stackrel{f_2}{\longrightarrow} \mathbb C^{N_2} \stackrel{f_3}{\longrightarrow} \mathbb C^{2N_3} \stackrel{f_4}{\longrightarrow} \mathbb C^{N'_0} \longrightarrow 0.
\end{equation}
Here $N'_0$ is the number of \emph{inner} vertices in~$M$, while $N_2$~-- the number of all 2-faces, and $N_3$~-- the number of all tetrahedra.

We assume that all vertices in~$M$ are numbered from~$1$ to their total number~$N_0$, and we ascribe ``coordinates'' $\zeta_1,\dots,\zeta_{N_0}$ to them. These are arbitrary complex numbers with the only condition
\begin{equation}\label{znez}
\zeta_i \ne \zeta_j \qquad \text{for} \quad i\ne j.
\end{equation}
We will also use notation
\begin{equation}\label{zij}
\zeta_{ij} \stackrel{\rm def}{=} \zeta_i - \zeta_j .
\end{equation}

\begin{remark}
The numbers~$k$ of mappings~$f_k$ in~\eqref{c} begin from~$2$ and not~$1$ in order to make them consistent with similar complexes that include two more mappings: $f_1$ on the left and $f_5$ on the right, see, e.g., \cite[formula~(5)]{bkm}. In this paper, however, we do not use complexes longer than~\eqref{c}.
\end{remark}

Both spaces $\mathbb C^{N'_0}$ in~\eqref{c} consist, by def\/inition, of column vectors whose components, denoted~$u_i$ for the left-hand space and~$v_i$ for the right-hand space, are in one-to-one correspondence with inner vertices~$i$. More formally, each of these spaces is a space over~$\mathbb C$ with inner vertices as its basis.

Spaces $\mathbb C^{N_2}$ and~$\mathbb C^{2N_3}$ are a bit more complicated. To explain them, we begin with two auxiliary spaces: $W_2$ whose basis is formed of all \emph{pairs}~$(s,i)$, where $s$ is a 2-face and $i\in s$~-- its vertex, and~$W_3$ whose basis is formed of all pairs~$(r,i)$, where $r$ is a tetrahedron and $i\in r$~-- its vertex. Thus, $\dim W_2=3N_2$ and $\dim W_3=4N_3$. We use notations like~$x_{s,i}$ or~$y_{r,i}$ for coordinates of a vector~$x\in W_2$ or~$y\in W_3$.

Then we introduce space~$V_2\subset W_2$ consisting of vectors whose coordinates obey
\begin{equation}\label{s}
\begin{split}
& x_{s,i}+x_{s,j}+x_{s,k}=0, \\
& \zeta_i x_{s,i}+\zeta_j x_{s,j}+\zeta_k x_{s,k}=0
\end{split}
\end{equation}
for every 2-face~$s$ with vertices $i$, $j$ and~$k$, and similarly space~$V_3\subset W_3$ consisting of vectors whose coordinates obey
\begin{equation}\label{r}
\begin{split}
& y_{r,i}+y_{r,j}+y_{r,k}+y_{r,\ell}=0, \\
&\zeta_i y_{r,i}+\zeta_j y_{r,j}+\zeta_k y_{r,k}+\zeta_{\ell} y_{r,\ell}=0
\end{split}
\end{equation}
for every tetrahedron~$r$ with vertices $i$, $j$, $k$ and~$\ell$.

Thus, a vector $x\in V_2$ is determined by specifying just one its coordinate in each 2-face~$s$, and assuming that $i<j<k$, we will take coordinate~$x_{s,i}$ for that. The space~$\mathbb C^{N_2}$ consists, by def\/inition, of column vectors whose coordinates are these~$x_{s,i}$ for all~$s$.

Similarly, a vector $y\in V_3$ is determined by specifying just two of its coordinates in each tetrahedron~$r$, and assuming that $i<j<k<\ell$, we will take $y_{r,i}$ and~$y_{r,j}$ for that. The space~$\mathbb C^{2N_3}$ consists, by def\/inition, of column vectors whose coordinates are these $y_{r,i}$ and~$y_{r,j}$ for all~$r$.

Linear mapping~$f_2$ makes, by def\/inition, the following~$x_{s,i}$ from given~$u_i$:
\begin{equation}\label{f2}
f_2\colon \  x_{s,i}=\big(\zeta_{ij}^{-1}-\zeta_{ik}^{-1}\big)u_i-\zeta_{ij}^{-1}u_j+\zeta_{ik}^{-1}u_k,
\end{equation}
where 2-face~$s$ has vertices $i<j<k$.

Linear mapping~$f_3$ makes, by def\/inition, the following~$y_{r,i}$ from given~$x_{s,i}$:
\begin{equation}\label{f3}
f_3\colon \
\begin{cases}
y_{r,i} = x_{(ijk),i}-x_{(ij\ell),i}+x_{(ik\ell),i} ,\\
y_{r,j} = x_{(ijk),j}-x_{(ij\ell),j}-x_{(jk\ell),j} ,
\end{cases}
\end{equation}
where tetrahedron~$r$ has vertices $i<j<k<\ell$, and by~$(ijk)$ and so on we denote the 2-faces of~$r$ containing the indicated vertices.

\begin{cnv}\label{cnv:o}
In general, when we write an $n$-simplex in this paper as $(i_0\dots i_n)$, the (numbers of) its vertices are ordered so that $i_0<\dots<i_n$, if other ordering is not indicated explicitly.
\end{cnv}

To def\/ine linear mapping~$f_4$, we must f\/ix an orientation of~$M$, i.e., a consistent orientation of all its tetrahedra. This results in ascribing a sign
\begin{equation}\label{eps}
\epsilon_r=\pm 1
\end{equation}
to each tetrahedron~$r$ with vertices $i<j<k<\ell$ in the following way: $\epsilon_r=1$ if the orientation of~$r$ determined by the order $i$, $j$, $k$, $\ell$ of vertices coincides with the mentioned consistent orientation, and $\epsilon_r=-1$ otherwise. Mapping~$f_4$ makes, by def\/inition, the following~$v_i$ from given~$y_{r,i}$:
\begin{equation}\label{f4}
f_4\colon \  v_i=\sum_{r\ni i} \epsilon_r y_{r,i},
\end{equation}
the sum goes, of course, over all tetrahedra containing vertex~$i$.

\begin{theorem}\label{th:c}
The chain~\eqref{c} of vector spaces and linear mappings defined as above is indeed a~chain complex, i.e.,
\[
f_4\circ f_3=0,\qquad f_3\circ f_2=0.
\]
\end{theorem}

\begin{proof}
Theorem~\ref{th:c} can be proved by direct calculations. For a conceptual explanation of the origin of~\eqref{c}, see~\cite[Subsection~3.2]{bk}.
\end{proof}

\subsubsection{Invariants from Reidemeister torsions}\label{sss:R}

We want to calculate some Reidemeister torsions for chain complex~\eqref{c}. The complex~\eqref{c} as it is will, however, never be acyclic for a manifold with non-empty boundary. This is because its algebraic Euler characteristic is $-N_2+2N_3 \ne 0$. To be more exact, in the case of non-empty boundary $N_2>2N_3>N'_2$, where $N'_2$ is the number of \emph{inner} 2-faces.

Actually, this allows us to introduce not one but many torsions. First, we take an \emph{ordered} subset~$\mathcal C$ of boundary faces of cardinality $\# \mathcal C=2N_3-N'_2$. Second, we consider, instead of~$\mathbb C^{N_2}$, its subspace~$(V_2)_{\mathcal C}$ consisting of those vectors whose coordinates corresponding to boundary faces outside~$\mathcal C$ are zero. We assume also that the coordinates corresponding to inner edges go f\/irst and are ordered in the same f\/ixed way for all~$\mathcal C$, and then go the coordinates belonging to~$\mathcal C$ and ordered also as~$\mathcal C$. Third, we def\/ine a new complex~-- a subcomplex of~\eqref{c}~-- by replacing $\mathbb C^{N_2}$ with~$(V_2)_{\mathcal C}$ and restricting naturally linear mappings $f_2$ and~$f_3$~-- just taking their submatrices corresponding to~$(V_2)_{\mathcal C}$.

It can be checked that Theorem~\ref{th:c} remains valid for this new chain complex corresponding to the set~$\mathcal C$, and we def\/ine its Reidemeister torsion\footnote{This construction can be also interpreted in terms of torsions for chain complexes with nonvanishing homologies, see~\cite[Subsection~3.1]{turaev}. We leave this as an exercise for the reader, just mentioning that a subset~$\mathcal C$ determines a basis in the homology space corresponding to the middle term of the complex \emph{conjugate} to~\eqref{c}, i.e., with arrows reversed and matrices $f_2$, $f_3$, $f_4$ transposed.} in a standard way as
\begin{equation}\label{tau}
\tau_{\mathcal C} = \frac{(\minor f_3)_{\mathcal C}}{\minor f_2\,\minor f_4},
\end{equation}
where the minors are chosen according to the rules for a matrix $\tau$-chain, see~\cite[Subsection~2.1]{turaev}. Moreover, we can take the the same minors of both $f_2$ and~$f_4$ for all~$\mathcal C$, and this is ref\/lected in~\eqref{tau} by writing the subscript~$\mathcal C$ only at~$(\minor f_3)$.

\begin{remark}
Of course, for some~$\mathcal C$'s, both $(\minor f_3)_{\mathcal C}$ and~$\tau_{\mathcal C}$ will vanish, or, speaking more strictly, there will be no $\tau$-chain.
\end{remark}

Now we introduce the following quantities, where the letter~$I$ stays for ``invariant'', and the superscript~$(0)$ is to emphasize that these are our ``old'' invariants, to be deformed soon:
\begin{equation}\label{I0}
I_{\mathcal C}^{(0)} = \frac{\displaystyle\prod_{\substack{{\rm inner}\\ {\rm 2\text{-}faces\ }s}} \zeta_{s_2s_3}}{\displaystyle\prod_{\substack{{\rm inner}\\ {\rm edges\ }\ell}} \zeta_{\ell_1\ell_2}\prod_{\substack{{\rm all}\\ {\rm tetrahedra\ }r}} \zeta_{r_3r_4}} \cdot \tau_{\mathcal C},
\end{equation}
where we use the following notations:
\begin{itemize}\itemsep=0pt
\item $\ell_1$ and~$\ell_2$ are the vertices of an inner edge~$\ell$ taken in the increasing order: $\ell_1<\ell_2$,
\item similarly, $s_1<s_2<s_3$ are the vertices of an inner 2-face~$s$, and
\item $r_1<r_2<r_3<r_4$~-- the vertices of a tetrahedron~$r$.
\end{itemize}

\begin{theorem}\label{th:iR}
The values~\eqref{I0} for all~$\mathcal C$ form a multicomponent invariant of manifold~$M$ with a~fixed boundary triangulation, defined up to an overall (the same for all~$\mathcal C$) sign.
\end{theorem}

\begin{proof}
To prove that $I_{\mathcal C}^{(0)}$, for a given~$\mathcal C$, is a manifold invariant, it is enough to prove its invariance under:
\begin{enumerate}\itemsep=0pt
\item\label{i:co} a change of order of \emph{inner} vertices,
\item\label{i:23} a Pachner move $2\leftrightarrow 3$,
\item\label{i:14} a Pachner move $1\leftrightarrow 4$.
\end{enumerate}

For items \ref{i:23} and~\ref{i:14}, we refer the reader to \cite[Theorem~4]{bk}, where this is proved in a more general situation\footnote{In the formulation of~\cite[Theorem~4]{bk}, the boundary~$\partial M$ is assumed to be one-component. This, however, is not used in the proof. The point is that, actually, \eqref{iR} vanishes for \emph{multi}component~$\partial M$. In this paper we, nevertheless, do not put away the multicomponent case, because our aim is to introduce a deformation of~\eqref{iR} which may behave dif\/ferently.}.

To prove \ref{i:co}\footnote{It must be admitted that the (more general) analogue of~\ref{i:co} should have been proven also already in~\cite{bk}.}, we note that a change of vertex order implies the corresponding change of bases in spaces $V_2$ and~$V_3$. To see the change of Reidemeister torsion, we must calculate determinants of transition matrices between bases in $V_2$ and~$V_3$, or their inverses~-- ratios between exterior products of all new and all old coordinates in the corresponding space:
\[
\frac{\bigwedge (x_{s,i})_{\rm new}}{\bigwedge (x_{s,i})_{\rm old}} \qquad \text{and} \qquad \frac{\bigwedge (y_{r,i})_{\rm new}}{\bigwedge (y_{r,i})_{\rm old}}.
\]
As one can deduce from \eqref{s} and~\eqref{r} such relations as, for instance,
\[
\frac{x_{s,j}}{x_{s,i}}=-\frac{\zeta_{ik}}{\zeta_{jk}} \qquad \text{and} \qquad \frac{y_{r,k}\wedge y_{r,\ell}}{y_{r,i}\wedge y_{r,j}}=\frac{\zeta_{ij}}{\zeta_{k\ell}},
\]
it is not hard to check that the invariance of~\eqref{I0} really holds.

As for the sign of each~$I_{\mathcal C}^{(0)}$, it is not determined uniquely because of arbitrariness of ordering basis vectors in our vector spaces. It can be easily seen, however, that any change in such ordering makes the same ef\/fect on the sign of \emph{every}~$I_{\mathcal C}^{(0)}$: the only basis vectors that dif\/fer in two complexes corresponding to two~$\mathcal C$'s belong to these~$\mathcal C$'s, and their order is f\/ixed because the~$\mathcal C$'s are ordered. Also, any possible sign ambiguities in the above transformations \ref{i:co}, \ref{i:23} and~\ref{i:14} af\/fect the signs of all~$I_{\mathcal C}^{(0)}$ in the same way. This proves that all~$I_{\mathcal C}^{(0)}$ are determined up to one overall sign.
\end{proof}

\subsubsection{Invariants made from Reidemeister torsions in terms of Grassmann algebra}\label{sss:Rw}

We put in correspondence to each unoriented 2-face~$s$ in the triangulation a Grassmann gene\-ra\-tor~$a_s$, and to each unoriented\footnote{The orientations of tetrahedra are actually important for us, but we will take them into account in another way, see Def\/inition~\ref{d:vt}.} tetrahedron~$r$ two Grassmann generators $b_r^{(1)}$ and~$b_r^{(2)}$.

We denote $\mathbf a$ the column vector made of all~$a_j$, and $\mathbf b$ the column vector made of all $b_r^{(1)}$ and~$b_r^{(2)}$.

\begin{definition}\label{d:t}
For a tetrahedron~$r$, we introduce its \emph{Grassmann weight} as follows:
\begin{equation}\label{tw}
W_r = \exp \Phi_r ,
\end{equation}
where
\begin{equation}\label{Phi}
\Phi_r =
\begin{pmatrix}
b_r^{(1)} & b_r^{(2)}
\end{pmatrix}
\begin{pmatrix}
1 & -1 & 1 & 0 \\
-\zeta_{r_2r_3}^{-1}\zeta_{r_1r_3} & \zeta_{r_2r_4}^{-1}\zeta_{r_1r_4} & 0 & -1
\end{pmatrix}
\begin{pmatrix}
a_{(r_1r_2r_3)}\\ a_{(r_1r_2r_4)}\\ a_{(r_1r_3r_4)}\\ a_{(r_2r_3r_4)}
\end{pmatrix} .
\end{equation}
In~\eqref{Phi}, $r_1$, $r_2$, $r_3$, $r_4$ are the vertices of~$r$ in the increasing order; $(r_1r_2r_3)$ and the like are the faces of~$r$ having corresponding vertices. The $2\times 4$ matrix in the r.h.s.\ is a block of which matrix~$f_3$ is built, in accordance with \eqref{f3}, \eqref{s} and~\eqref{r}.
\end{definition}

\begin{definition}\label{d:vf}
For every inner vertex~$i$, we introduce the following \emph{vertex-face differential operator} in our Grassmann algebra:
\[
d_i^{\mathbf a} = \sum_{\textrm{2-faces }s\ni i}d_{s,i}^{\mathbf a},
\]
where
\[
d_{s,i}^{\mathbf a}=\begin{cases}
(1/\zeta_{s_1s_2}-1/\zeta_{s_1s_3})\partial/\partial a_s & \text{if }i=s_1, \\
(-1/\zeta_{s_1s_2})\partial/\partial a_s & \text{if }i=s_2, \\
(1/\zeta_{s_1s_3})\partial/\partial a_s & \text{if }i=s_3;
\end{cases}
\]
$s_1$, $s_2$, $s_3$ are the vertices of~$s$ in the increasing order. The coef\/f\/icients at~$\partial/\partial a_s$ are matrix elements of matrix~$f_2$, in accordance with~\eqref{f2}.
\end{definition}

\begin{definition}\label{d:vt}
Also, we introduce one more operator for every inner vertex~$i$, the \emph{vertex-tetrahedron differential operator}:
\[
d_i^{\mathbf b}=\sum_{\textrm{tetrahedra }r\ni i}d_{r,i}^{\mathbf b},
\]
where
\[
d_{r,i}^{\mathbf b}=\begin{cases}
\epsilon_r\,\partial/\partial b_r^{(1)} & \text{if }i=r_1, \\
\epsilon_r\,\partial/\partial b_r^{(2)} & \text{if }i=r_2, \\
\epsilon_r\bigl(-(\zeta_{r_1r_4}/\zeta_{r_3r_4})\partial/\partial b_r^{(1)}-(\zeta_{r_2r_4}/\zeta_{r_3r_4})\partial/\partial b_r^{(2)}\bigr) & \text{if }i=r_3, \\
\epsilon_r\bigl(\zeta_{r_1r_3}/\zeta_{r_3r_4}\partial/\partial b_r^{(1)}+\zeta_{r_2r_3}/\zeta_{r_3r_4}\partial/\partial b_r^{(2)}\bigr) & \text{if }i=r_4;
\end{cases}
\]
$r_1$, $r_2$, $r_3$, $r_4$ are the vertices of~$r$ in the increasing order. The coef\/f\/icients at $\partial/\partial b_r^{(1)}$ and~$\partial/\partial b_r^{(2)}$ are matrix elements of matrix~$f_4$, in accordance with \eqref{f4} and~\eqref{r}.
\end{definition}

\begin{cnv}\label{cnv:d}
Let $h$ be a homogeneous element of Grassmann algebra of degree~$m$, and~$d$ a~homogeneous Grassmann dif\/ferential operator of degree~$n$. Then $d^{-1}h$ means \emph{any homogeneous} element~$f$ of Grassmann algebra (of degree $(m+n)$, of course) such that $df=h$.
\end{cnv}

\begin{theorem}\label{th:GR}
The following function of Grassmann variables~$a_s$ living on boundary $2$-faces:
\begin{equation}\label{gfR}
\mathbf T =
\idotsint\prod_{\substack{{\rm all}\\ {\rm tetrahedra\ }r}} W_r \cdot
\Biggl(\prod_{\substack{{\rm inner}\\ {\rm  vertices\ }i}} d_i^{\mathbf a}\Biggr)^{-1}1 \cdot
\Biggl(\prod_{\substack{{\rm inner}\\ {\rm  vertices\ }i}} d_i^{\mathbf b}\Biggr)^{-1}1 \cdot
\mathrm d\mathbf b\,\mathrm d\mathbf a_{\rm inner }
\end{equation}
is the generating function for torsions~$\tau_{\mathcal C}$, see~\eqref{tau}, in the sense that
\[
\mathbf T = \sum_{\mathcal C} \tau_{\mathcal C} \prod_{s\in\mathcal C} a_s .
\]
In~\eqref{gfR}, $\mathrm d\mathbf a_{\rm inner }$ and $\mathrm d\mathbf b$ stay for the products
\[
\mathrm d\mathbf a_{\rm inner } =\prod_{\substack{{\rm inner}\\ {\rm  2\text{-}faces\ }s}} \mathrm da_s, \qquad
\mathrm d\mathbf b =\prod_{\substack{{\rm all}\\ {\rm tetrahedra\ }r}} \mathrm db_r^{(1)}\,\mathrm db_r^{(2)},
\]
and the expressions $({\rm dif\/ferential\ operator})^{-1}1$ are defined according to Convention~{\rm \ref{cnv:d}}.
\end{theorem}

\begin{proof}
It is always possible to choose both $\displaystyle\Biggl(\prod_{\substack{{\rm inner}\\ {\rm  vertices\ }i}} d_i^{\mathbf a}\Biggr)^{-1}1$ and $\displaystyle\Biggl(\prod_{\substack{{\rm inner}\\ {\rm  vertices\ }i}} d_i^{\mathbf b}\Biggr)^{-1}1$ as Grassmann \emph{monomials}~-- products of some Grassmann generators and a numeric factor. Then it can be seen that the numeric factor is exactly $(\minor f_2)^{-1}$ or~$(\minor f_4)^{-1}$ respectively (compare formula~\eqref{tau}), where the rows in $\minor f_2$ or the columns in $\minor f_4$ correspond to the mentioned Grassmann generators. Then it is not hard to deduce that the factor at~$\displaystyle \prod_{s\in\mathcal C} a_s$ in~$\mathbf T$ is nothing but~$\tau_{\mathcal C}$, and in passing we see that it does not depend on the choice of monomials. Nor will it change, of course, if we take a linear combination of monomials satisfying the same Grassmann dif\/ferential equation
\begin{equation*}
({\rm dif\/ferential\ operator})f = 1 .\tag*{\qed}
\end{equation*}
\renewcommand{\qed}{}
\end{proof}

\begin{definition}\label{d:gfR}
We call the following function of Grassmann variables~$a_s$ corresponding to bounda\-ry 2-faces~$s$ \emph{generating function} of invariants~$I_{\mathcal C}^{(0)}$:
\begin{equation}\label{iR}
\mathbf F = \sum_{\mathcal C} I_{\mathcal C}^{(0)} \prod_{s\in \mathcal C} a_s.
\end{equation}
\end{definition}

It follows from Theorem~\ref{th:GR} and formula~\eqref{I0} that
\begin{equation}\label{F}
\mathbf F = \frac{\displaystyle\prod_{\substack{{\rm inner}\\ {\rm 2\text{-}faces\ }s}} \zeta_{s_2s_3}}{\displaystyle\prod_{\substack{{\rm inner}\\ {\rm edges\ }\ell}} \zeta_{\ell_1\ell_2}\prod_{\substack{{\rm all}\\ {\rm tetrahedra\ }r}} \zeta_{r_3r_4}} \cdot \mathbf T .
\end{equation}

\subsubsection[Opening the way to generalizations: Grassmann algebra relations corresponding to Pachner moves]{Opening the way to generalizations:\\ Grassmann algebra relations corresponding to Pachner moves}\label{sss:um}

Writing the multicomponent invariants~$\mathbf F$~\eqref{F} for the l.h.s.\ and r.h.s.\ of the Pachner move $2\to 3$, see Fig.~\ref{f:23}, we come to the following relation in Grassmann algebra:
\begin{equation}\label{23_0}
\frac{\zeta_{23}}{\zeta_{34}\zeta_{35}}\int \mathcal W_{1234} \mathcal W_{1235} \,\mathrm da_{123} =
-\frac{1}{\zeta_{45}}\iiint \mathcal W_{1245} \mathcal W_{1345} \mathcal W_{2345} \,\mathrm da_{145}\,\mathrm da_{245}\,\mathrm da_{345},
\end{equation}
where we def\/ine
\begin{equation}\label{cW_0}
\mathcal W_r = \iint W_r \,\mathrm db_r^{(1)} \,\mathrm db_r^{(2)} ,
\end{equation}
and where we have also, of course, checked the sign separately. We also write $\mathcal W_{1234}$, $a_{123}$ and so on instead of more pedantic $\mathcal W_{(1234)}$ and~$a_{(123)}$.

The Grassmann function~$\mathcal W_r$ can be treated as the invariant~$\mathbf F$ for just one tetrahedron~$r$.

It is quite clear that Grassmann algebra relations like~\eqref{23_0} can be taken themselves as a~starting point for building manifold invariants and a topological quantum f\/ield theory, and there is, in principle, \emph{no need for the components of~$\mathcal W_r$ to be related to torsions} of any chain complex. This is the idea that we are going to explore.

\subsection{The deformed relations}\label{ss:tildeW}

Our search for new Grassmann algebra relations associated with Pachner moves started with ``deformations'' of the weight~$\mathcal W_r$: what terms (if any) can be added to $\mathcal W_r$ so that relation~\eqref{23_0} stays valid? Of course we have not yet found \emph{all} possible deformations; the miracle is, however, that such deformations \emph{do exist}, as shown by the results of our search, presented below.

\subsubsection{Deformation of degree~0}\label{sss:d0}

We introduce the following deformation of the weight~$\mathcal W_r$ for a tetrahedron~$r=(r_1r_2r_3r_4)$ belonging to an oriented triangulated PL manifold~$M$ with boundary:
\begin{equation}\label{calW}
\tilde{\mathcal W}_r \stackrel{\rm def}{=} \mathcal W_r + \epsilon_r \zeta_{r_3r_4} \alpha_r ,
\end{equation}
where $\epsilon_r$ is the tetrahedron orientation (see formula~\eqref{eps} and explanation after it), and $\alpha_r$ is some even element in the Grassmann algebra, depending on the tetrahedron~$r$. Thus, $\alpha_r$ commutes with any other element, for instance, $\alpha_r$ may be a scalar${}\in\mathbb F$, and this possibility looks (at this moment) the most natural. So, assuming $\alpha_r \in \mathbb F$, the term $\epsilon_r \zeta_{r_3r_4} \alpha_r$ in~\eqref{calW} has the degree~$0$, and we call $\tilde{\mathcal W}_r$ \emph{deformation of degree~$0$} of~$\mathcal W_r$.

\begin{remark}\label{rem:zeta}
Both factors $\epsilon_r$ and~$\zeta_{r_3r_4}$ are introduced in formula~\eqref{calW} because such def\/inition of~$\alpha_r$ clarif\/ies formulas below, like \eqref{alpha}--\eqref{alpha45}.
\end{remark}

The direct substitution of $\tilde{\mathcal W}_r$, given by the ansatz~\eqref{calW}, in place of $\mathcal W_r$ in~\eqref{23_0}, using GAP~\cite{GAP} and our package PL~\cite{PL} and assisted also by maxima~\cite{maxima}, gives the following result: \eqref{23_0} holds for~$\tilde{\mathcal W}_r$, provided the $\alpha$'s in the r.h.s.\ of Pachner moves are expressed through the $\alpha$'s in the l.h.s.\ as follows:
\begin{equation}\label{alpha}
\begin{split}
& \zeta_{35}\alpha_{1235} - \zeta_{34}\alpha_{1234} = \zeta_{45}\alpha_{1245},\\
& \zeta_{25}\alpha_{1235} - \zeta_{24}\alpha_{1234} = \zeta_{45}\alpha_{1345},\\
& \zeta_{15}\alpha_{1235} - \zeta_{14}\alpha_{1234} = \zeta_{45}\alpha_{2345}.
\end{split}
\end{equation}
Note that we have written the $\alpha$'s corresponding to the r.h.s.\ of Pachner move also in the \emph{r.h.s.} of~\eqref{alpha}, and we of course assume the same orientations for both sides of the move.

Equations~\eqref{alpha} can be written in the following elegant form. First, it will be convenient for us to write~$\alpha_r$ also as~$\alpha_{\{ijkl\}}$, where $i$, $j$, $k$ and~$l$ are the vertices of~$r$ taken in an \emph{arbitrary} order (recall that, according to Convention~\ref{cnv:o}, we usually assume $i<j<k<l$ when we write $r=(ijkl)$). Now, choose an oriented edge~$(kl)$; given also a consistent orientation of all tetrahedra, the link of~$(kl)$ can be considered as made of \emph{oriented} edges~$(ij)$.

In an oriented edge~$(ij)$, the order of vertices $i$ and~$j$ determines its orientation, so this is an \textbf{exception} where Convention~\ref{cnv:o} does not work!

\begin{definition}\label{d:alpha}
We say that a \emph{consistent system} of~$\alpha$'s is given for a Pachner move $2\leftrightarrow 3$ or $1\leftrightarrow 4$, if
\begin{equation}\label{alpha3}
\sum_{\substack{(ij)\text{ in the}\\ \text{oriented link of }(kl)\\ \text{in the l.h.s.}}} \zeta_{ij}\, \alpha_{\{ijkl\}} = \sum_{\substack{(ij)\text{ in the}\\ \text{oriented link of }(kl)\\ \text{in the r.h.s.}}} \zeta_{ij}\, \alpha_{\{ijkl\}} \,.
\end{equation}
If $(kl)$ is present in only one of the sides of Pachner move, then of course the sum in the corresponding side of~\eqref{alpha3} is zero. We also say that a consistent system of~$\alpha$'s is given for any triangulated oriented manifold~$M$ with boundary, if
\begin{equation}\label{alphaM}
\sum_{\substack{(ij)\text{ in the}\\ \text{oriented link of }(kl)}} \zeta_{ij}\, \alpha_{\{ijkl\}} = 0
\end{equation}
for any inner (non-boundary) edge $(kl)\subset M$.
\end{definition}

The three equations~\eqref{alpha} have already the form~\eqref{alpha3} for edges $12$, $13$ and~$23$ respectively; other relations follow from these, like
\begin{equation}\label{alpha45}
0 = \zeta_{12}\alpha_{1245}-\zeta_{13}\alpha_{1345}+\zeta_{23}\alpha_{2345}
\end{equation}
for edge~$45$.

\begin{definition}\label{d:tildeW}
Let a consistent system of $\alpha$'s be given for a Pachner move $2\leftrightarrow 3$ or $1\leftrightarrow 4$, then for a tetrahedron~$r=(r_1r_2r_3r_4)$, taking part in the move, we introduce its \emph{deformed Grassmann weight}
\begin{equation}\label{tildeW}
\tilde W_r = \exp\big(\Phi_r + \epsilon_r \zeta_{r_3r_4} \alpha_r b_r^{(2)} b_r^{(1)} \big).
\end{equation}
Here $\Phi_r$ is the same  as before, see formula~\eqref{Phi}.
\end{definition}

Obviously, the analogue of \eqref{cW_0} holds:
\begin{equation}\label{cW}
\tilde{\mathcal W}_r = \iint \tilde W_r \,\mathrm db_r^{(1)} \,\mathrm db_r^{(2)} .
\end{equation}

\begin{remark}\label{r:gauges}
The weight~$\tilde W_r$ introduced in formula~\eqref{tildeW} coincides, essentially, with the second solution of pentagon equation in~\cite{g}, with the constant multiplier~$\mu$ omitted and in a dif\/ferent gauge. See~\cite[Subsection~5.3]{bk}, where the transition from one gauge to the other is described for the case of ``undeformed'' weights~-- and this transition stays the same in the ``deformed'' case.
\end{remark}

\begin{theorem}\label{th:tildeW}
The weights defined according to \eqref{tildeW} and~\eqref{cW} satisfy the following $2\to 3$ relation:
\begin{equation}\label{23}
\frac{\zeta_{23}}{\zeta_{34}\zeta_{35}}\int \tilde{\mathcal W}_{1234} \tilde{\mathcal W}_{1235} \,\mathrm da_{123} =
-\frac{1}{\zeta_{45}}\iiint \tilde{\mathcal W}_{1245} \tilde{\mathcal W}_{1345} \tilde{\mathcal W}_{2345} \,\mathrm da_{145}\,\mathrm da_{245}\,\mathrm da_{345},
\end{equation}
if the $\alpha$'s form a consistent system for this move.

Moreover, the weights defined according to the same formulas, together with the operators $d_i^{\mathbf a}$ and~$d_i^{\mathbf b}$ given in Definitions {\rm \ref{d:vf}} and~{\rm \ref{d:vt}}, satisfy the following $1\to 4$ relation:
\begin{gather}
\frac{1}{\zeta_{34}} \tilde{\mathcal W}_{1234} = - \frac{1}{\zeta_{15}\zeta_{45}}
\idotsint \tilde W_{1235} \tilde W_{1245} \tilde W_{1345} \tilde W_{2345} \cdot (d_5^{\mathbf a})^{-1}1 \cdot (d_5^{\mathbf b})^{-1}1 \nonumber\\
\hphantom{\frac{1}{\zeta_{34}} \tilde{\mathcal W}_{1234} =}{} \times \mathrm db_{1235}^{(1)} \, \mathrm db_{1235}^{(2)} \, \mathrm db_{1245}^{(1)} \, \mathrm db_{1245}^{(2)} \, \mathrm db_{1345}^{(1)} \, \mathrm db_{1345}^{(2)} \, \mathrm db_{2345}^{(1)} \, \mathrm db_{2345}^{(2)} \nonumber\\
\hphantom{\frac{1}{\zeta_{34}} \tilde{\mathcal W}_{1234} =}{}
\times \mathrm da_{125}\,\mathrm da_{135}\,\mathrm da_{145}\,\mathrm da_{235}\,\mathrm da_{245}\,\mathrm da_{345}  ,\label{14}
\end{gather}
if the $\alpha$'s form a consistent system for this move.
\end{theorem}

\begin{proof}
We have already explained the formula~\eqref{23}. Similarly, formula~\eqref{14} has been checked on a computer using our package~PL~\cite{PL}.
\end{proof}

\begin{definition}\label{d:uw}
We denote $(d_5^{\mathbf a})^{-1}1=u_5$ and $(d_5^{\mathbf b})^{-1}1=w_5$ and call them \emph{vertex weights} for vertex~$5$; similarly below $u_i$ and~$w_i$ for any vertex~$i$. Each of them can be chosen as a monomial of degree one: $u_i$ containing Grassmann generator~$a_s$ for some 2-face $s\ni i$, and $w_i$ containing $b_r^{(1\text{ or }2)}$ for some tetrahedron $r\ni i$. If, additionally, $r\supset s$, we refer to both $u_i$ and~$w_i$ as \emph{corresponding to tetrahedron~$r$}.
\end{definition}

\subsubsection{Deformation of degree~4}\label{sss:d4}

It turns out that there exists also another deformation $\tilde{\mathcal W}_r$ of our weight~$\mathcal W_r$, where, instead of the term of degree~$0$, a  term of degree~$4$ is added to~$\mathcal W_r$:
\begin{equation}\label{calWdeg4}
\tilde{\mathcal W}_r \stackrel{\rm def}{=} \mathcal W_r + \epsilon_r \zeta_{r_3r_4} c_{r_1r_2r_3r_4} a_{r_1r_2r_3}  a_{r_1r_2r_4} a_{r_1r_3r_4} a_{r_2r_3r_4} ,
\end{equation}
where
\[
c_{r_1r_2r_3r_4} = \prod_{1\le i<j\le 4} \zeta_{r_ir_j} ,
\]
and $\epsilon_r=\pm 1$ is the same as in Subsection~\ref{sss:d0}. For instance, in the following Theorem~\ref{th:deg4}, $\epsilon_{(1234)}=\epsilon_{(1345)}=1$ and $\epsilon_{(1235)}=\epsilon_{(1245)}=\epsilon_{(2345)}=-1$.

The detailed study of the weight~\eqref{calWdeg4} is still in progress; here we just report about the result which we formulate as the following theorem.

\begin{theorem}\label{th:deg4}
Tetrahedron Grassmann weights~$\tilde{\mathcal W}_r$ defined according to~\eqref{calWdeg4} satisfy the same $2\leftrightarrow 3$ relation~\eqref{23}.
\end{theorem}

\begin{proof}
Direct calculation.
\end{proof}

\begin{remark}
The weight~\eqref{calWdeg4} appeared f\/irst in formula~(11) of preprint~\cite{g}, but in a dif\/ferent gauge, see again Remark~\ref{r:gauges}.
\end{remark}

\section[Four dimensions: deformed relations $3\to 3$ and $2\leftrightarrow 4$]{Four dimensions: deformed relations $\boldsymbol{3\to 3}$ and $\boldsymbol{2\leftrightarrow 4}$}\label{s:4d}

\subsection{Recalling the undeformed relations}\label{ss:recalling4d}

\subsubsection{The chain complex}\label{sss:c4}

In four dimensions, the starting point for introducing the ``undeformed'' 4-simplex weight is again an exotic chain complex~-- a four-dimensional analogue of complex~\eqref{c}. We write out here its simplif\/ied~-- short~-- version, suited for studying Pachner moves $3\to 3$ and $2\leftrightarrow 4$. This will be enough for explaining then the current results of our symbolic calculations; a longer version of the complex will be presented in a separate paper~\cite{ks}.

We consider a triangulated orientable four-manifold~$M$ with boundary and with the additional requirement that the triangulation has \emph{no inner vertices}. Our short chain complex for such a~manifold\footnote{Of course, complex~\eqref{c4} can be written out also for a triangulation having inner vertices, but in such case~\eqref{c4} is not enough to obtain nonzero torsions.} is:
\begin{equation}\label{c4}
0 \longrightarrow \mathbb C^{N'_2} \stackrel{f_3}{\longrightarrow} \mathbb C^{2N_3} \stackrel{f_4}{\longrightarrow} \mathbb C^{3N_4} \longrightarrow 0.
\end{equation}
Here $N'_2$ is the number of \emph{inner} 2-faces, while $N_3$ and $N_4$ are the numbers of \emph{all} 3-faces and 4-simplices in~$M$, respectively.

As before, all triangulation vertices have numbers~$i$ from~$1$ to~$N_0$ and complex ``coordinates''~$\zeta_i$ with the condition~\eqref{znez}, and we again use notation~\eqref{zij} for their dif\/ferences.

\begin{remark}
Again, like in the three-dimensional case, the numbering of mappings in~\eqref{c4} begins from~$f_3$ because notations $f_1$ and~$f_2$ (and actually $f_5$ and~$f_6$) are reserved for longer complexes.
\end{remark}

The spaces in~\eqref{c4} are much like those in~\eqref{c}. Again, we begin with auxiliary spaces: $W_2$ whose basis is formed of all pairs~$(s,i)$, where $s$ is now an inner 2-face and $i\in s$~-- its vertex, $W_3$ whose basis is formed of all pairs~$(r,i)$, where $r$ is a 3-face and $i\in r$~-- its vertex, and $W_4$ whose basis is formed of all pairs~$(u,i)$, where $u$ is a 4-simplex and $i\in u$~-- its vertex. Thus, $\dim W_2=3N_2'$, $\dim W_3=4N_3$, and $\dim W_4=5N_4$. We use notations $x_{s,i}$, $y_{r,i}$ or~$z_{u,i}$ for coordinates of a vector~$x\in W_2$, $y\in W_3$ or~$z\in W_4$ respectively.

Then we introduce spaces $V_2\subset W_2$ and $V_3\subset W_3$ consisting of vectors whose coordinates obey the same relations \eqref{s} and~\eqref{r} as in the three-dimensional case, and the space $V_4\subset W_4$ consisting of vectors~$z$ whose coordinates also obey similar relations:
\begin{equation*}
\begin{split}
& z_{u,i}+z_{u,j}+z_{u,k}+z_{u,\ell}+z_{u,m}=0, \\
& \zeta_i z_{u,i}+\zeta_j z_{u,j}+\zeta_k z_{u,k}+\zeta_{\ell} z_{u,\ell}+\zeta_m z_{u,m}=0 \end{split}
\end{equation*}
for every 4-simplex~$u$ with vertices $i$, $j$, $k$, $\ell$ and~$m$.

In the same way as in the three-dimensional case, we take for a vector $x\in V_2$ one coordinate in each (inner) 2-face~$s$, namely coordinate corresponding to the vertex with the smallest number, to identify $V_2$ with~$\mathbb C^{N_2'}$, and similarly identify $V_3$ with~$\mathbb C^{2N_3}$ and $V_4$ with~$\mathbb C^{3N_4}$, taking two coordinates in each 3-face or three coordinates in each 4-simplex, respectively.

Linear mapping~$f_3$ is def\/ined by the old formulas~\eqref{f3}, and linear mapping~$f_4$ is also def\/ined in a similar way. Namely, it makes the following $z_{u,i}$ from given~$y_{r,i}$:
\begin{equation*}
f_4\colon \
\begin{cases}
z_{u,i} = y_{(ijk\ell),i}-y_{(ijkm),i}+y_{(ij\ell m),i}-y_{(ik\ell m),i}\,,\\
z_{u,j} = y_{(ijk\ell),j}-y_{(ijkm),j}+y_{(ij\ell m),j}+y_{(jk\ell m),j}\,,\\
z_{u,k} = y_{(ijk\ell),k}-y_{(ijkm),k}-y_{(ik\ell m),k}+y_{(jk\ell m),k}\,,
\end{cases}
\end{equation*}
where 4-simplex~$u$ has vertices $i<j<k<\ell<m$, and by~$(ijk\ell)$ and so on we denote the 3-faces of~$u$ containing the indicated vertices.

\begin{theorem}\label{th:c4}
The chain~\eqref{c4} of vector spaces and linear mappings defined as above is indeed a~chain complex, i.e.,
\[
f_4\circ f_3=0.
\]
\end{theorem}

\begin{proof}
Direct calculation.
\end{proof}

\subsubsection{Changing the gauge}\label{sss:gauge}

Before constructing a Grassmann weight out of matrix~$f_4$, we see it convenient to make a~``gauge transformation'' for this matrix. Namely, we multiply each column corresponding to each 3-face~$ijk\ell$ by~$\zeta_{k\ell}$ (matrix~$f_4$ has, of course, two columns for each 3-face). We denote the resulting matrix~$\tilde f_4$.

Accordingly, we \emph{divide} each \emph{row} of matrix~$f_3$ by~$\zeta_{k\ell}$, if it corresponds to a 3-face~$ijk\ell$. Additionally, we multiply each column of~$f_3$ by~$\zeta_{jk}$ if it corresponds to an inner 2-face~$ijk$. The obtained matrix is denoted~$\tilde f_3$. Obviously, the chain complex condition $\tilde f_4 \tilde f_3 = 0$ still holds.

Below, Def\/initions \ref{d:W4} and~\ref{d:w4} show how to construct Grassmann weights in order to interpret the torsions of complex~\eqref{c4} (in the new gauge) in terms of Grassmann algebra. Namely, in formula~\eqref{W4} for a 4-simplex weight, the three expressions in parentheses have matrix elements of the three rows of~$\tilde f_4$ as coef\/f\/icients at Grassmann variables (while multiplier $1/\zeta_{45}$ is just for elegance, see Remark~\ref{r:9}), and in formulas~\eqref{2fops4d}, the coef\/f\/icients at partial derivatives are matrix elements of~$\tilde f_3$.

\subsubsection{The formulas for weights}\label{sss:udw4}

We present now the undeformed 4-simplex weight, and the dif\/ferential operators corresponding to 2-faces, in the gauge described above\footnote{They have been already written in such form in preprint~\cite{4}; we only change the letter $W$ to $\mathcal W$, because our weight~\eqref{W4} is an analogue of~\eqref{cW_0} rather than~\eqref{tw}. The situation with gauges in four dimensions is largely the same as in three dimensions, compare Remark~\ref{r:gauges}.}. Instead of writing out the Grassmann weight~$\mathcal W_{ijklm}$ for a general 4-sim\-plex $(ijklm)$, we write out just~$\mathcal W_{12345}$ for readability; $\mathcal W_{ijklm}$ is obtained by the obvious substitution $1\mapsto i,\dots,5\mapsto m$.
\begin{definition}\label{d:W4}
The undeformed Grassmann weight corresponding to 4-sim\-plex $(12345)$ is the following function of Grassmann variables $a_{i_1i_2i_3i_4}$ and~$b_{i_1i_2i_3i_4}$ attached to each tetrahedron~$(i_1i_2i_3i_4)$~-- a 3-face of~$(12345)$:
\begin{gather}
\mathcal W_{12345} \stackrel{\rm def}{=} \frac{1}{\zeta_{45}}
(\zeta_{34}a_{1234}-\zeta_{35}a_{1235}+\zeta_{45}a_{1245}-\zeta_{45}a_{1345})\nonumber \\
\hphantom{\mathcal W_{12345} \stackrel{\rm def}{=}}{} \times (\zeta_{34}b_{1234}-\zeta_{35}b_{1235}+\zeta_{45}b_{1245}+\zeta_{45}a_{2345}) \nonumber\\
\hphantom{\mathcal W_{12345} \stackrel{\rm def}{=}}{} \times (-\zeta_{14}a_{1234}-\zeta_{24}b_{1234}+\zeta_{15}a_{1235}+\zeta_{25}b_{1235}-\zeta_{45}b_{1345}+\zeta_{45}b_{2345}) .\label{W4}
\end{gather}
\end{definition}

\begin{remark}\label{r:9}
The factor $1 / \zeta_{45}$ in~\eqref{W4} cancels out in all monomials obtained after expanding~\eqref{W4}. So the weight~\eqref{W4} is bilinear in~$\zeta$'s and, moreover, the coef\/f\/icient at each product of $a$'s and/or $b$'s has the form $\zeta_{ij}\zeta_{kl}$, where the subscripts may coincide. The total number of such terms (monomials with nonzero coef\/f\/icients~$\zeta_{ij}\zeta_{kl}$) in the expansion of~\eqref{W4} is~72, see~\cite[Appendix]{4}.
\end{remark}

\begin{definition}\label{d:w4}
For a 2-face~$s=(ijk)$, we introduce the dif\/ferential operator~$d_{ijk}$ as
\[
d_{ijk} = \sum_{{\rm tetrahedra\ }t\supset s} d_{t,s} ,
\]
where $d_{t,s}$ are the following operators, which we again prefer to write out putting numbers rather than letters in subscripts: we take tetrahedron~$t=(1234)$ and its four faces, having in mind that, for an arbitrary tetrahedron~$(ijkl)$ (remember that $i<j<k<l$), the substitution $1\mapsto i,\dots,4\mapsto l$ must be done. So, the operators are:
\begin{equation}\label{2fops4d}
d_{(1234),s}= \begin{cases}
(\zeta_{23}/\zeta_{34})\,\partial/\partial a_{1234} - (\zeta_{13}/\zeta_{34})\,\partial/\partial b_{1234} & {\rm if\ } s=(123),\\
-(\zeta_{24}/\zeta_{34})\,\partial/\partial a_{1234} + (\zeta_{14}/\zeta_{34})\,\partial/\partial b_{1234} & {\rm if\ } s=(124),\\
\partial/\partial a_{1234} & {\rm if\ } s=(134),\\
-\partial/\partial b_{1234} & {\rm if\ } s=(234).
\end{cases}
\end{equation}
\end{definition}

\subsection[The deformed four-simplex weight and the relations $3\to 3$ and $2\leftrightarrow 4$]{The deformed four-simplex weight and the relations $\boldsymbol{3\to 3}$ and $\boldsymbol{2\leftrightarrow 4}$}\label{ss:rels4d}

Our deformed four-simplex Grassmann weight~-- an analogue of~\eqref{calW}~-- is introduced as follows. First, we need a ``constant'' (not depending on any simplices) odd element of Grassmann algebra, we denote it~$e$. Then, we need an even element~$\alpha_{\{ijklm\}}$ for every four-simplex~$(ijklm)$, satisfying an analogue of relation~\eqref{alpha3}, namely relation~\eqref{alpha4} below. The curly brackets around the subscripts of~$\alpha$ mean the following: in order to write~\eqref{alpha4} in a simple way, it will be convenient for us to assume that the~$\alpha$'s depend on an \emph{unordered} quintuple~$\{i,j,k,l,m\}=\{j,i,k,l,m\}=\dots=\{m,l,k,j,i\}$, that is, an~$\alpha$ does not change under a permutation of its indices.

Let there be a 2-face~$(klm)$ in the l.h.s.\ or/and r.h.s.\ of a Pachner move. Then, given a~consistent orientation (the same in both sides) of all 4-simplices, the link(s) of~$(klm)$ can be considered as made of \emph{oriented} edges~$(ij)$. Recall that oriented edges form an exception from Convention~\ref{cnv:o}!

\begin{definition}\label{d:W4d}
The deformed 4-simplex weight, entering in a $3\to 3$ or $2\leftrightarrow 4$ Pachner move, is
\begin{equation}\label{W4d}
\tilde {\mathcal W}_{ijklm} = \mathcal W_{ijklm} + \alpha_{\{ijklm\}} e ,
\end{equation}
where the $\alpha$'s are even elements of the Grassmann algebra satisfying the following relations for all 2-faces~$(klm)$
\begin{equation}\label{alpha4}
\sum_{\substack{(ij)\text{ in the}\\ \text{oriented link of }klm\\ \text{in the l.h.s.}}} \zeta_{ij}\, \alpha_{\{ijklm\}} = \sum_{\substack{(ij)\text{ in the}\\ \text{oriented link of }klm\\ \text{in the r.h.s.}}} \zeta_{ij}\, \alpha_{\{ijklm\}} \,.
\end{equation}
If $(klm)$ is present in only one of the sides of Pachner move, then of course the sum in the corresponding side of~\eqref{alpha4} iz zero.
\end{definition}

\begin{remark}
Our formula~\eqref{W4d} reveals some dif\/ference from its three-dimensional analogue~\eqref{calW}. While the analogue of the factor $\zeta_{r_3r_4}$ is absent from~\eqref{W4d} simply because of a dif\/ferent gauge, the role of~$\epsilon_r$ is played in~\eqref{W4d} by an object of apparently dif\/ferent nature~-- odd Grassmann element~$e$. And the experimental fact that formulas \eqref{33} and~\eqref{24} below hold for our weight~$\tilde {\mathcal W}_{ijklm}$ def\/initely deserves further research.
\end{remark}

\subsubsection[Move $3\to 3$]{Move $\boldsymbol{3\to 3}$}\label{sss:33}

Here is the description of the move $3\to 3$: the cluster of three 4-simplices $(12345)$, $(12346)$ and~$(12356)$, of which we think as being in the ``l.h.s.'', is replaced by the cluster of simplices $(12456)$, $(13456)$ and~$(23456)$ in the ``r.h.s.''. The boundary of either side consists of tetrahedra $(1245)$, $(1246)$, $(1256)$, $(1345)$, $(1346)$, $(1356)$, $(2345)$, $(2346)$, and~$(2356)$. The \emph{inner} tetrahedra are, however, dif\/ferent: $(1234)$, $(1235)$ and~$(1236)$ in the l.h.s., and $(1456)$, $(2456)$ and~$(3456)$ in the r.h.s. Also, there is one inner 2-face~$(123)$ in the l.h.s., and one inner 2-face~$(456)$ in the r.h.s.

\begin{theorem}
\label{th:33}
The following identity, corresponding naturally to the $3\to 3$ Pachner move, holds:
\begin{gather}
\int \tilde {\mathcal W}_{12345} \tilde {\mathcal W}_{12346} \tilde {\mathcal W}_{12356} w_{123} \frac{\mathrm da_{1234}\,\mathrm db_{1234}}{\zeta_{34}} \frac{\mathrm da_{1235}\,\mathrm db_{1235}}{\zeta_{35}}\frac{\mathrm da_{1236}\,\mathrm db_{1236}}{\zeta_{36}} \nonumber\\
 \qquad{} = \int \tilde {\mathcal W}_{12456} \tilde {\mathcal W}_{13456} \tilde {\mathcal W}_{23456} w_{456} \frac{\mathrm da_{1456} \,\mathrm db_{1456}}{\zeta_{56}} \frac{\mathrm da_{2456} \,\mathrm db_{2456}}{\zeta_{56}} \frac{\mathrm da_{3456} \,\mathrm db_{3456}}{\zeta_{56}} ,\label{33}
\end{gather}
where
\[
w_{123} = d_{123}^{-1}1, \qquad w_{456} = d_{456}^{-1}1
\]
$($for instance, $w_{123} = \zeta_{23}^{-1}\zeta_{34}a_{1234}$ and $w_{456}=-b_{1456}$, recall Convention~{\rm \ref{cnv:d})}.
\end{theorem}

\begin{proof}
It can be checked that both sides of \eqref{33} cannot contain the~$\alpha$'s in the (total) degree more than one. The ``constant'' terms~-- not containing the~$\alpha$'s~-- form themselves the ``undeformed'' equation of preprint~\cite{4}, which can be, and has been, verif\/ied separately on a computer. Then, it is not hard to see that, in order to prove the equality between the terms linear in~$\alpha$'s, it is enough to do so for just two following sets of~$\alpha$'s:
\[
\alpha_{\{12345\}}=\alpha_{\{12346\}}=\alpha_{\{12356\}}=\alpha_{\{12456\}}=\alpha_{\{13456\}}=\alpha_{\{23456\}}=1,
\]
and
\begin{gather*}
\alpha_{\{12345\}}=\zeta_6,\qquad \alpha_{\{12346\}}=\zeta_5,\qquad \alpha_{\{12356\}}=\zeta_4,\\
\alpha_{\{12456\}}=\zeta_3,\qquad \alpha_{\{13456\}}=\zeta_2,\qquad \alpha_{\{23456\}}=\zeta_1.
\end{gather*}
This has also been done on a computer, using GAP~\cite{GAP} and our PL package~\cite{PL}.
\end{proof}

\subsubsection[Move $2\to 4$]{Move $\boldsymbol{2\to 4}$}\label{sss:24}

Next, we consider the following $2\to 4$ move: the cluster of two 4-simplices $(12345)$ and $(12346)$ is replaced by the cluster of four 4-simplices $(12356)$, $(12456)$, $(13456)$ and~$(23456)$. The bounda\-ry of both sides consists of tetrahedra $(1235)$, $(1236)$, $(1245)$, $(1246)$, $(1345)$, $(1346)$, $(2345)$ and~$(2346)$.

In the l.h.s., there is one inner tetrahedron~$(1234)$ and no inner 2-faces.

In the r.h.s., there are six inner tetrahedra $(1256)$, $(1356)$, $(1456)$, $(2356)$, $(2456)$ and~$(3456)$, and four inner 2-faces $(156)$, $(256)$, $(356)$ and~$(456)$. The $d$-operators for these 2-faces are, according to~\eqref{2fops4d}, as follows:
\begin{gather}
d_{156}=\frac{\partial}{\partial a_{1256}} + \frac{\partial}{\partial a_{1356}} + \frac{\partial}{\partial a_{1456}}  , \qquad
d_{256}= - \frac{\partial}{\partial b_{1256}} + \frac{\partial}{\partial a_{2356}} + \frac{\partial}{\partial a_{2456}}  , \nonumber\\
d_{356}= - \frac{\partial}{\partial b_{1356}} - \frac{\partial}{\partial b_{2356}} + \frac{\partial}{\partial a_{3456}}  , \qquad
d_{456}= - \frac{\partial}{\partial b_{1456}} - \frac{\partial}{\partial b_{2456}} - \frac{\partial}{\partial b_{3456}}  .
\label{dr24}
\end{gather}
So, one can check\footnote{Recalling again Convention~\ref{cnv:d}.} that, for instance,
\[
a_{1256}b_{1256}a_{3456}b_{3456}
\]
is suitable as $w_{156,256,356,456}$ in formula~\eqref{24} below.

\begin{cnj}
\label{cnj:24}
The following identity, corresponding naturally to the $2\to 4$ move, holds:
\begin{gather}
\int \tilde {\mathcal W}_{12345} \tilde {\mathcal W}_{12346} \frac{\mathrm da_{1234}\,\mathrm db_{1234}}{\zeta_{34}} \nonumber\\
 \qquad{} = - \zeta_{56} \int \tilde {\mathcal W}_{12356} \tilde {\mathcal W}_{12456} \tilde {\mathcal W}_{13456} \tilde {\mathcal W}_{23456} w_{156,256,356,456} \label{24}\\
\qquad{} \times \frac{\mathrm da_{1256} \,\mathrm db_{1256}}{\zeta_{56}} \frac{\mathrm da_{1356} \,\mathrm db_{1356}}{\zeta_{56}} \frac{\mathrm da_{1456} \,\mathrm db_{1456}}{\zeta_{56}} \frac{\mathrm da_{2356} \,\mathrm db_{2356}}{\zeta_{56}} \frac{\mathrm da_{2456} \,\mathrm db_{2456}}{\zeta_{56}} \frac{\mathrm da_{3456} \,\mathrm db_{3456}}{\zeta_{56}} .
 \nonumber
\end{gather}
\end{cnj}

Here, like in Theorem~\ref{th:33}, the weights~$\tilde{\mathcal W}$ in~\eqref{24} are taken for the 4-simplices which are respectively in the l.h.s.\ and r.h.s.; the weight~$w$ is taken for the inner 2-faces; and the integration is performed in $a$ and~$b$ corresponding to the inner tetrahedra.

\begin{proof}[Partial verif\/ication]
First, like in Theorem~\ref{th:33}, both sides of~\eqref{24} contain only terms of degree${}\le 1$ in all~$\alpha$'s, and the ``constant'' parts form themselves the ``undeformed'' equation of~\cite{4} that has been verif\/ied directly and fully. Second, the equality between the parts linear in~$\alpha$'s have been checked not in full, but for some arbitrarily chosen \emph{values} of~$\zeta$'s (assuming that we are working in the f\/ield $\mathbb F=\mathbb Q$ of rational numbers), such as
\begin{equation*}
\zeta_1=0,\qquad \zeta_2=1,\qquad \zeta_3=3,\qquad \zeta_4=8,\qquad \zeta_5=17,\qquad \zeta_6=21.\tag*{\qed}
\end{equation*}
\renewcommand{\qed}{}
\end{proof}

\begin{remark}
The factor~$(-\zeta_{56})$ before the integral in the r.h.s.\ of~\eqref{24} is naturally interpreted as corresponding to the \emph{inner edge}~$56$ in the cluster of four 4-simplices. There are of course no inner edges in any cluster of two or three 4-simplices considered in this paper.
\end{remark}

\begin{remark}
Is both formulas \eqref{33} and~\eqref{24}, specif\/ic orderings/numberings of vertices $1,\dots, 6$ were involved. For other orderings, similar formulas still hold, but we do not discuss it in this paper, where our modest aim is just to show the existence of such relations. Note that in a~simpler three-dimensional situation, necessary arguments about (change of) vertex ordering are given in the proof of Theorem~\ref{th:G}.
\end{remark}

\subsection[Conjectured invariant of moves $3\to 3$ and $2\leftrightarrow 4$]{Conjectured invariant of moves $\boldsymbol{3\to 3}$ and $\boldsymbol{2\leftrightarrow 4}$}
\label{ss:i}

Formulas \eqref{33} and~\eqref{24} lead to a conjectured invariant of moves $3\to 3$ and $2\leftrightarrow 4$. We would like to formulate this as the following Conjecture~\ref{cng:i4p}, even if this conjecture looks somewhat preliminary at this stage of research, mainly because the question of choosing interesting consistent systems of~$\alpha$'s remains open. Note, however, that one possibility to satisfy equations~\eqref{alpha4} is to take all $\alpha$'s simply equal to~$1$!

\begin{cnj}\label{cng:i4p}
The following expression, written for an arbitrary triangulated $4$-manifold with boundary, remains invariant under moves $3\to 3$ and $2\leftrightarrow 4$:
\begin{equation}\label{ti}
\pm \prod_{\substack{\text{\rm over inner}\\ \text{\rm edges }ij } }
\, \int \!\!\!\!\! \prod_{\substack{\text{\rm over all}\\ \text{\rm 4-simplices }ijklm } } \!\!\! \tilde {\mathcal W}_{ijklm}
\cdot w \cdot
\!\!\! \prod_{\substack{\text{\rm over inner}\\ \text{\rm tetrahedra }ijkl } } \!\!\! \frac{\rm \mathrm da_{ijkl}\, \mathrm db_{ijkl}}{\zeta_{kl}}\, ,
\end{equation}
where
\begin{equation}\label{tii}
w= \Biggl( \prod_{\substack{\text{\rm over inner}\\ \text{\rm 2-faces }ijk } } d_{ijk} \Biggr)^{-1} 1 .
\end{equation}
\end{cnj}

The sign~$\pm$ in~\eqref{ti} corresponds to the fact that we did not specify the exact order in the products; most likely, there exists some elegant formula relating this order and this sign.

\section[Again three dimensions: definition of a deformed invariant and nontriviality check]{Again three dimensions: def\/inition of a deformed invariant\\ and nontriviality check}\label{s:di}

\subsection{Some combinatorics before introducing the deformed invariant}\label{ss:comb}

\begin{definition}\label{d:o}
Let there be a f\/inite set~$X$ and its cover $\{S_i,\; i=1,\dots, N\}$ consisting of its subsets~$S_i\subset X$. Let there be also given an injection~$f$ choosing an element in each of the covering sets:
\[
f\colon\; S_i\mapsto r_i\in S_i, \qquad r_i\ne r_j \quad \text{if} \quad i\ne j.
\]
We call the mapping~$f$ \emph{orderly} if the following relation generates correctly a \emph{partial order} for sets~$S_i$:
\[
S_i \le S_j \qquad \text{if} \quad f(S_i)\in S_j.
\]
\end{definition}

Clearly, an orderly mapping~$f$ sends $S_k$ into an element not belonging to any preceding set $S_i<S_k$. We denote
\[
R_k^{(f)} = S_k \setminus (\text{sets preceding }S_k).
\]

\begin{definition}
An \emph{elementary move} is the following change of an orderly mapping~$f$: change one image~$f(S_k)$ to some other element~$\tilde r_k\in R_k^{(f)}$; other images $f(S_i)$, $i\ne k$, remain intact.
\end{definition}

\begin{remark}
An elementary move may change the partial order on sets~$S_i$, and thus, generally speaking, \emph{new} subsets $R_i^{(h)}\subset S_i$ appear as its result, where $h$ denotes the result of applying the above elementary move to~$f$. It does not, however, change~$R_k^{(f)}=R_k^{(h)}$, because it does not af\/fect the order of sets preceding~$S_k$. So, the inverse to elementary move is also elementary move.
\end{remark}

\begin{lemma}\label{l:o}
For a given cover $\{S_i,\; i=1,\dots, N\}$ of a finite set~$X$, any orderly mapping~$f$ can be transformed into any other orderly mapping~$g$ by a sequence of elementary moves.
\end{lemma}

\begin{proof}
Induction in~$N$~-- the number of sets~$S_i$. For $N=1$, the theorem obviously holds.

For arbitrary~$N$, denote~$S_m$ a \emph{maximal}~$S_i$, with respect to the partial order determined by mapping~$f$. Then $R_m^{(f)}\subset S_m$ is its part not belonging to \emph{any} other~$S_i$. Consider the set $X'=X\setminus R_m^{(f)}$, its cover of cardinality $N-1$ consisting of all~$S_i$ except~$S_m$, and restrictions of $f$ and~$g$ on this cover. Due to induction hypothesis, the restriction of~$f$ can be transformed into the restriction of~$g$ by elementary moves. This means also that $f$ can be transformed into the mapping~$\tilde g$ which is by def\/inition the same as~$g$ except for, maybe, just one image,~$\tilde g(S_m)\stackrel{\rm def}{=}f(S_m)$.

Finally, changing $g$ to~$\tilde g$ is also an elementary move, because $R_m^{(g)}\supset R_m^{(f)}$, so $g(S_m)$ is changed within~$R_m^{(g)}$.
\end{proof}

\subsection[The deformed multicomponent invariant of a three-manifold with triangulated boundary]{The deformed multicomponent invariant of a three-manifold\\ with triangulated boundary}\label{ss:di}

For an inner vertex~$i$ in a triangulated orientable three-manifold~$M$ with nonempty bounda\-ry~$\partial M$, let $S_i$ be the set containing all tetrahedra in the star of~$i$ as its elements, and $X$ be the union of all~$S_i$. We are going to choose a tetrahedron in each~$S_i$ using an orderly mapping~$f$ in the sense of Def\/inition~\ref{d:o}.

We can construct~$f$ as follows: let~$r$ be any tetrahedron in the triangulation having at least one boundary (i.e., such that there is no tetrahedron on its other side) 2-face. Then we remove~$r$ from the triangulation. If, as a result of this, a vertex~$i$ uncloses~-- ceases to be inner, we put $f(S_i)=r$. Then we repeat this step until there remain no inner vertices.

\begin{remark}
If $S_i$ precedes $S_j$ in the sense of Def\/inition~\ref{d:o}, then vertex~$i$ uncloses in this process \emph{after} vertex~$j$.
\end{remark}

\begin{remark}\label{r:P}
Let there also be a marked PL ball~$B$ made of some tetrahedra in the triangulation, containing no inner vertices, in the sense that all its vertices are in~$\partial B$. Then the above construction of~$f$ can obviously be done without removing tetrahedra in~$B$. We say then that $f$ \emph{avoids} tetrahedra in~$B$.
\end{remark}

\begin{definition}
We say that we have \emph{chosen vertex weights according to orderly mapping~$f$} if, for each inner vertex~$i$, we have chosen $u_i$ and~$w_i$ corresponding to the tetrahedron~$f(S_i)$, see Def\/inition~\ref{d:uw}.
\end{definition}

\begin{theorem}\label{th:G}
The following function of Grassmann variables and coordinates~$\zeta_i$, considered up to an overall sign, is an invariant of a three-dimensional manifold~$M$ with a fixed triangulation of its boundary~$\partial M$:
\begin{gather}
\mathbf G =\frac{\displaystyle\prod_{\substack{{\rm inner}\\ {\rm 2\text{-}faces\ }s}} \zeta_{s_2s_3}}{\displaystyle\prod_{\substack{{\rm inner}\\ {\rm edges\ }\ell}} \zeta_{\ell_1\ell_2}\prod_{\substack{{\rm all}\\ {\rm tetrahedra\ }r}} \zeta_{r_3r_4}} \cdot \idotsint \exp( \mathbf b^{\mathrm T}f_3\mathbf a + \mathbf b^{\mathrm T}C\mathbf b )\nonumber\\
\phantom{\mathbf G =}{}\times \prod_{\substack{{\rm inner}\\ {\rm vertices\ }i}} u_i \cdot
\prod_{\substack{{\rm inner}\\ {\rm vertices\ }i}} w_i \cdot
\mathrm d\mathbf b \,\mathrm d\mathbf a  .\label{G}
\end{gather}
Here matrix~$f_3$ is the same as in Subsection~{\rm \ref{ss:C}}, and matrix~$C$ is made of blocks
\begin{equation}\label{G2}
\epsilon_r \begin{pmatrix} 0 & \zeta_{r_3r_4} \\ -\zeta_{r_3r_4} & 0 \end{pmatrix} ,
\end{equation}
where both rows and columns correspond to $b_r^{(1)}$ and~$b_r^{(2)}$, in this order; vertex weights $u_i$ and~$w_i$ are chosen according to any orderly mapping~$f$.
\end{theorem}

\begin{remark}
Both Grassmann variables and coordinates~$\zeta_i$ on which $\mathbf G$ depends belong, of course, to the boundary~$\partial M$.
\end{remark}

First, we prove the following lemma.

\begin{lemma}\label{l:e}
Elementary moves, changing the orderly mapping~$f$, do not affect the function~$\mathbf G$ defined according to~\eqref{G}.
\end{lemma}

\begin{proof}
Consider an elementary move changing~$f(S_k)$. Consider, in the situation \emph{before} the move, the chain of all~$S_i$ preceding~$f(S_k)$, and let $S_{i_1}$ be the minimal element in this chain. If there are more than 4 tetrahedra\footnote{With our def\/inition of triangulation, there might be also just 2 tetrahedra in the star; we leave this easy case to the reader.} in~$S_{i_1}$, we, using a suitable sequence of Pachner moves $2\to 3$ within the star of~$i_1$ (each lessening the number of tetrahedra in the actual star of~$i_1$ by one) and the f\/inal move $4\to 1$, remove the vertex~$i_1$ from triangulation; as there is exactly one vertex weight~$u_{i_1}$ and one~$w_{i_1}$ within the star, it can be seen that the moves can be chosen in such way that formula~\eqref{23} can be applied to all moves $2\to 3$ and formula~\eqref{14}~-- to the f\/inal move. Then we do the same for the second element~$S_{i_1}$ and so on, and f\/inally we do this for~$S_k$ itself.

The same can obviously be done for the situation \emph{after} the move, with the very same f\/inal triangulation. So, there is a chain of transformations connecting the two choices of~$f$ and not af\/fecting~$\mathbf G$.
\end{proof}

\begin{proof}[Proof of Theorem~\ref{th:G}]
Any triangulation of the interior of~$M$ can be transformed into any other triangulation by a sequence of relative Pachner moves, i.e., moves leaving the boundary triangulation intact. As this has been explained in detail in~\cite[Section~2]{dkm}, here we only note that, although the boundary in~\cite{dkm} was just a specially triangulated torus, the techniques generalize directly to the case of a general boundary.

We want now to apply Theorem~\ref{th:tildeW} to each of these Pachner moves. First, comparing \eqref{G} and~\eqref{G2} with~\eqref{tildeW}, we see that this will work if \emph{all} $\alpha$'s in Theorem~\ref{th:tildeW} are chosen to equal~$-2$ (condition~\ref{alpha3} is then obviously satisf\/ied). Second, we must ensure the right number of multi\-pliers~$u_i$ living on 2-faces and~$w_i$ living in tetrahedra involved in the move, namely zero for moves $2\leftrightarrow 3$ and $1\to 4$, and one~$u_5$ and one~$w_5$ for $4\to 1$, in accordance with \eqref{23} and~\eqref{14}. This is, however, easy to do having in mind Remark~\ref{r:P}: for all moves except $4\to 1$, construct $f$ avoiding all tetrahedra that are to be replaced, and for $4\to 1$~-- avoiding three of them. The possible change of~$f$ is, of course, justif\/ied by Lemmas \ref{l:o} and~\ref{l:e}.

And third, formulas \eqref{23} and~\eqref{14} may seem to require a specif\/ic ordering of vertices involved in any Pachner move. This question is solved by re-formulating our results in the ``symmetric'' gauge of~\cite{g}, in which the Grassmann weight of a tetrahedron cannot depend on the order of vertices in any way except its overall sign (recall again that the transition from one gauge to the other is described in~\cite[Subsection~5.3]{bk}).
\end{proof}

\begin{remark}
We see thus that the following consistent system of $\alpha$'s corresponds to inva\-riant~\eqref{G}:
\[
\alpha_r=-2 \quad \text{for all tetrahedra } r .
\]
It is of course interesting how to introduce an invariant with more general $\alpha$'s than just all the same constant. We leave this question for further work.
\end{remark}

\subsection[Example calculation: triangulated lenses without tubular neighborhoods of unknots]{Example calculation: triangulated lenses\\ without tubular neighborhoods of unknots}\label{ss:L}

\begin{figure}[t]
\centering
\includegraphics[scale=0.3]{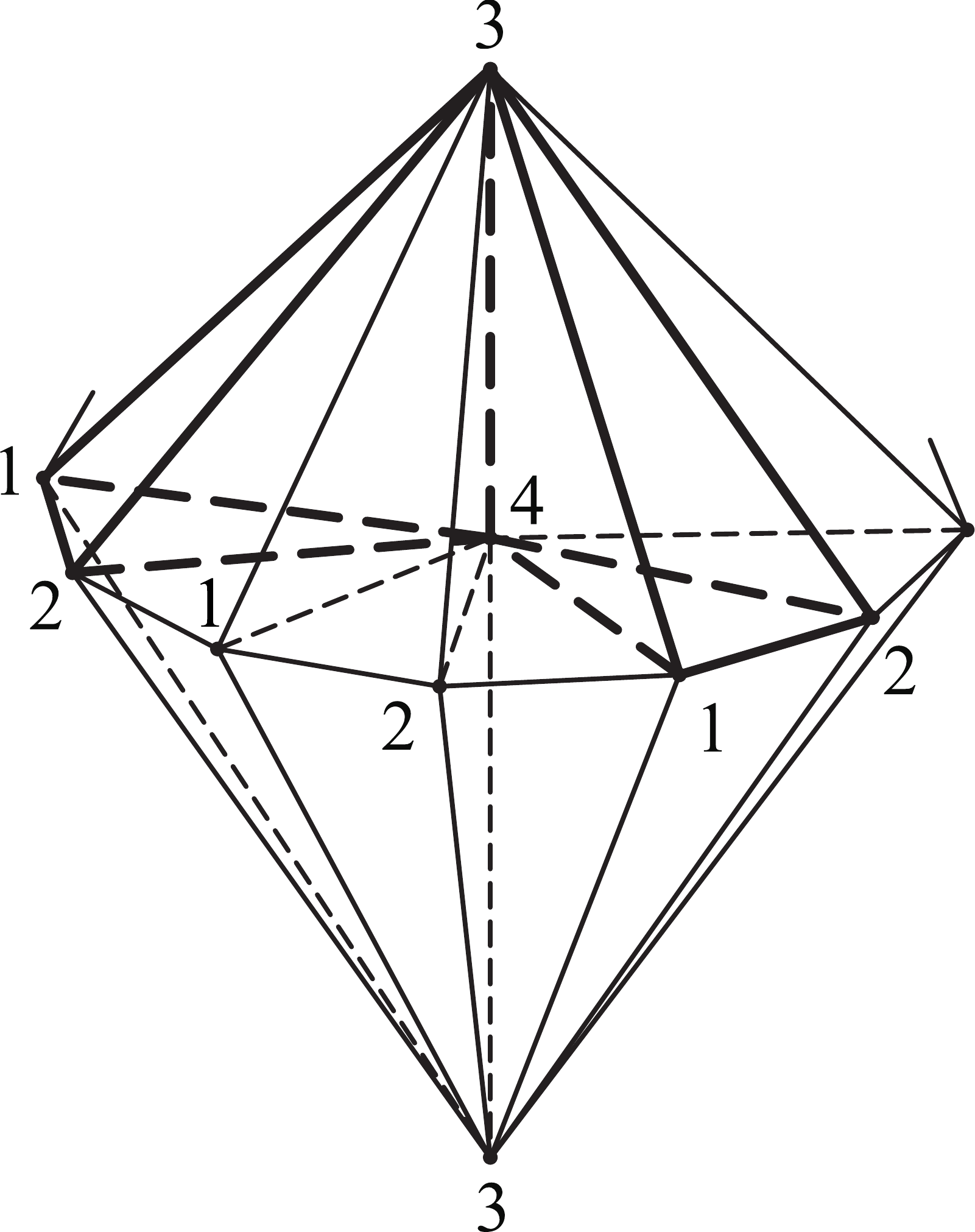}
\caption{Triangulated lens space with a chain of two tetrahedra.}
\label{f:L}
\end{figure}

Consider a three-dimensional lens space~$L(p,q)$, represented in a standard form of a bipyramid whose upper half-surface is glued to its lower half-surface after a rotation through angle~$2\pi q/p$ around the vertical axis (see, for instance, textbook~\cite{MF}). Let then the bipyramid be divided into~$4p$ tetrahedra, all with the same vertices~1, 2, 3, 4. A fragment of such triangulation is shown in Fig.~\ref{f:L}.
Then we choose an integer $n\ne 0 \mod p$, and two tetrahedra in this triangulation obtained one from another by a rotation through angle~$2\pi n/p$. In Fig.~\ref{f:L}, these are shown in boldface lines, for the case $n=2$.

\begin{figure}[t]
\centering
\includegraphics[scale=0.3]{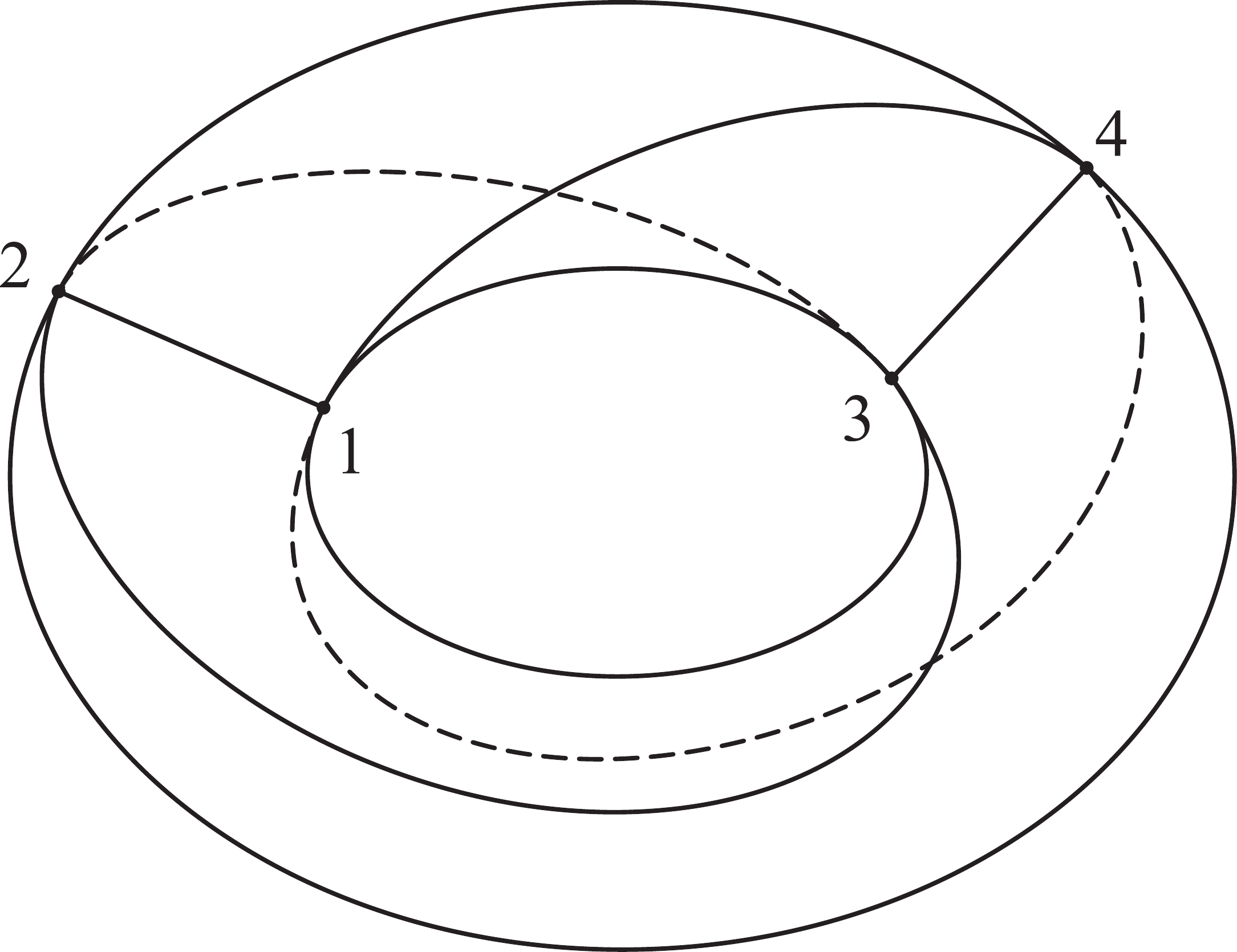}
\caption{The chain of two tetrahedra.}
\label{f:1a}
\end{figure}

\begin{figure}[t]
\centering
\includegraphics[scale=0.3]{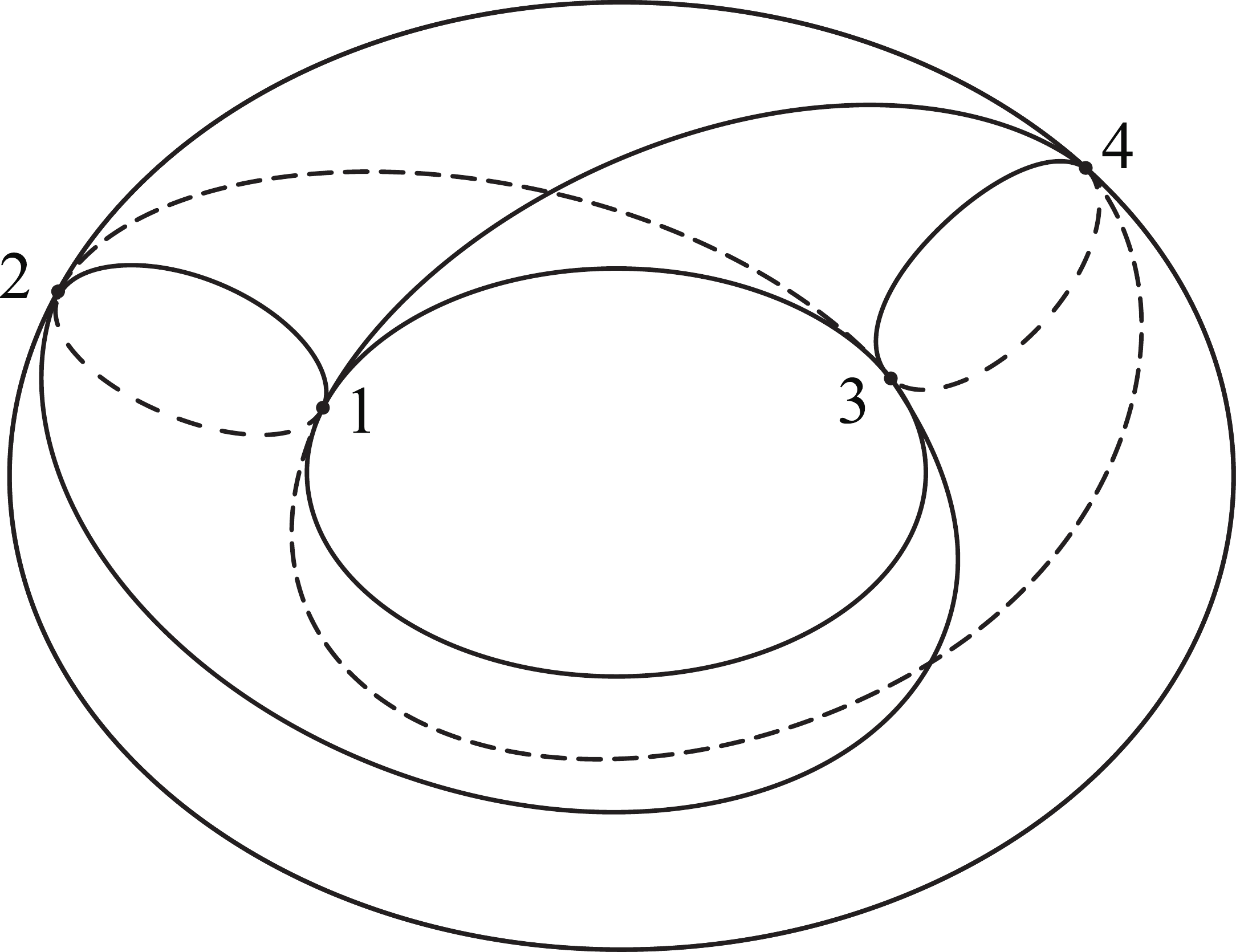}
\caption{After doubling edges $12$ and~$34$, the boundary of what remains of lens space becomes a torus triangulated in 8 triangles.}
\label{f:1d}
\end{figure}

We thus obtain a \emph{chain of two tetrahedra} in~$L(p,q)$, of the form pictured in Fig.~\ref{f:1a}.
Removing the interiors of these tetrahedra from~$L(p,q)$, and doubling common for two tetrahedra edges~$12$ and~$34$ as shown in Fig.~\ref{f:1d},
we obtain the lens space \emph{without a tubular neighborhood of an unknot} representing a 1-cycle determined by the number~$n$ above. Here an ``unknot'' in a closed 3-manifold~$M$ is characterized by the property that it can be represented by an unknotted line when $M$ is represented as a 3-ball with its surface glued in a proper way to itself. We denote the obtained 3-manifold with triangulated torus boundary as~$\tilde L$.

Next, we can calculate the invariant function~$\mathbf G$~\eqref{G} for~$\tilde L$, denoted~$\mathbf G_{\tilde L}$. As there are no inner vertices in our triangulation, \eqref{G} reduces to
\begin{equation}\label{GtildeL}
\mathbf G_{\tilde L} =\frac{\displaystyle\prod_{\substack{{\rm inner}\\ {\rm 2\text{-}faces\ }s}} \zeta_{s_2s_3}}{\displaystyle\prod_{\substack{{\rm inner}\\ {\rm edges\ }\ell}} \zeta_{\ell_1\ell_2}\prod_{\substack{{\rm all}\\ {\rm tetrahedra\ }r}} \zeta_{r_3r_4}} \cdot \idotsint \exp( \mathbf b^{\mathrm T}f_3\mathbf a + \mathbf b^{\mathrm T}C\mathbf b ) \, \mathrm d\mathbf b \,\mathrm d\mathbf a   .
\end{equation}
Our modest aim in this paper is just to show that our deformed invariants are nontrivial, that is, take some interesting values\footnote{Even for all $\alpha=1$.} that, we hope, deserve further investigation. We think that, at this stage, it is enough to present the results for the monomial in~$\mathbf G_{\tilde L}$ of degree zero in anticommuting variables; we denote it~$\mathrm G_{\tilde L}$. The essential point with~$\mathrm G_{\tilde L}$ is that its calculation really involves the deformation, that is, it would vanish if we took all $\alpha$'s equal to zero.

We simply present the following tables of directly calculated values of~$\mathrm G_{\tilde L}$. We made use of the fact that the integral in~\eqref{GtildeL} is the Pfaf\/f\/ian of the quadratic form in the exponent. Also, it was enough for us to do calculations for specif\/ic values of~$\zeta$'s using the already cited GAP system and our package~PL, although it makes little doubt that general formulas for a Pfaf\/f\/ian with a regular structure can be derived.
\[L(7,1)\colon\qquad
\begin{array}{c|c|c}
n & \zeta_1{=}1,\  \zeta_2{=}2,\  \zeta_3{=}3,\  \zeta_4{=}4 & \zeta_1{=}1,\  \zeta_2{=}2,\  \zeta_3{=}4,\  \zeta_4{=}3 \\ \hline
1 & 153 & 92 \\ \hline
2 & 313 & 324 \\ \hline
3 & 381 & 452
\end{array}
\]
\[L(7,2)\colon\qquad
\begin{array}{c|c|c}
n & \zeta_1{=}1,\  \zeta_2{=}2,\  \zeta_3{=}3,\  \zeta_4{=}4 & \zeta_1{=}1,\  \zeta_2{=}2,\  \zeta_3{=}4,\  \zeta_4{=}3 \\ \hline
1 & 12 & 61 \\ \hline
2 & 108 & 39 \\ \hline
3 & 153 & 92
\end{array}
\]
\[L(7,3)\colon\qquad
\begin{array}{c|c|c}
n & \zeta_1{=}1,\  \zeta_2{=}2,\  \zeta_3{=}3,\  \zeta_4{=}4 & \zeta_1{=}1,\  \zeta_2{=}2,\  \zeta_3{=}4,\  \zeta_4{=}3 \\ \hline
1 & 39 & 108 \\ \hline
2 & 92 & 153 \\ \hline
3 & 61 & 12
\end{array}
\]

\begin{remark}
Recall that every value of~$\mathrm G_{\tilde L}$ in these three tables is def\/ined up to a sign.
\end{remark}

\begin{remark}
$L(7,3)$ was here, of course, just for controle, as it is known to be homeomorphic to~$L(7,2)$ and, moreover, a PL homeomophism can be described explicitly in a simple and direct way as disassembling a bipyramid representing one of these spaces into a set of tetrahedra and then assembling them back into another bipyramid, see, e.g., textbook~\cite{MF}.
\end{remark}

\section{Discussion}\label{s:D}

Our hope is that our Pacher-move-like algebraic relations will lead to constructing new topological quantum f\/ield theories (TQFT's). As they will be based on the Grassmann--Berezin calculus, they can be called ``fermionic TQFT's''. This is, of course, especially interesting in four dimensions.

It would be interesting to relate our research to ideas of the paper~\cite{BNG}. More specif\/ically, it is interesting to know whether the formulas in~\cite{BNG} can be extended/deformed to incorporate a~case outside of a pure Reidemeister torsion, as well as four-dimensional manifolds.

On the other side, we considered in the present paper only the case where a ``coordinate''~$\zeta_i$ is put in correspondence to a vertex~$i$, and not a more general case where a universal cover of the manifold is involved and dif\/ferent coordinates correspond to dif\/ferent lifts of a vertex. In terms of~\cite{BNG}, this corresponds to considering a trivial f\/lat connection and a trivial bundle over the manifold. So, one direction of our further research may involve our ``deformed'' Pacher-move-like algebraic relations together with nontrivial f\/lat connections. In the ``undeformed'' case and a~slightly dif\/ferent specif\/ic theory, this was done in papers~\cite{M1,M2}.

One more very important generalization may be made through considering general consistent systems of~$\alpha$'s (see \eqref{alpha3} and~\eqref{alpha4}). This is expected to be related to some interesting homological problems. Also, some parameter counting suggests that in the four-dimensional case, interesting deformations of \emph{degree~$1$ in variables $a$ and~$b$} are expected to exist, which may be even more interesting than the deformation~\eqref{W4d}.

Finally, we emphasize that our calculations in Subsection~\ref{ss:L} simply show the nontriviality of invariants arising from our ``deformed'' relations. The nature of these invariants and possible relations to existing theories are still to be investigated, and again especially in four dimensions.

\subsection*{Acknowledgements}

This paper has been written with partial f\/inancial support from Russian Foundation for Basic Research, Grant no.~10-01-00088-a, and a grant from the Academic Senate of Moscow State University of Instrument Engineering and Computer Sciences. I also thank the referees for their constructive and helpful comments.

\pdfbookmark[1]{References}{ref}
\LastPageEnding

\end{document}